\documentclass[preprint]{aastex6}
\usepackage{color}

\newcommand{\itbold}[1]{\textbf{\textit{#1}}}
\newcommand{\pdif}[2]{\frac{\partial #1}{\partial #2}}
\newcommand{\odif}[2]{\frac{d #1}{d #2}}

\begin{document}

\title{Multidimensional Vlasov–Poisson Simulations with High-order
Monotonicity- and Positivity-preserving Schemes}

\author{Satoshi Tanaka\altaffilmark{1}, Kohji Yoshikawa}
\affil{Center for Computational Sciences, University of Tsukuba, 1-1-1
Tennodai, Tsukuba, Ibaraki 305-8577, Japan}

\author{Takashi Minoshima}
\affil{Department of Mathematical Science and Advanced Technology, Japan
Agency for Marine-Earth Science and Technology, 3173-25, Syowa-machi,
Kanazawa-ku, Yokohama, Kanagawa 236-0001, Japan}

\and
\author{Naoki Yoshida}
\affil{Department of Physics, The University of Tokyo, Bunkyo, Tokyo 113-0033, Japan}
\affil{Kavli Institute for the Physics and Mathematics of the Universe,
The University of Tokyo, Kashiwa, Chiba 277-8583, Japan}

\altaffiltext{1}{stanaka@ccs.tsukuba.ac.jp}

\begin{abstract}
 We develop new numerical schemes for Vlasov--Poisson equations with
 high-order accuracy. Our methods are based on a spatially
 monotonicity-preserving (MP) scheme and are modified suitably so that
 positivity of the distribution function is also preserved.  We adopt an
 efficient semi-Lagrangian time integration scheme that is more
 accurate and computationally less expensive than the three-stage TVD
 Runge-Kutta integration. We apply our spatially fifth- and
 seventh-order schemes to a suite of simulations of collisionless
 self-gravitating systems and electrostatic plasma simulations,
 including linear and nonlinear Landau damping in one dimension and
 Vlasov--Poisson simulations in a six-dimensional phase space.  The
 high-order schemes achieve a significantly improved accuracy in
 comparison with the third-order positive-flux-conserved scheme adopted
 in our previous study.  With the semi-Lagrangian time integration, the
 computational cost of our high-order schemes does not significantly
 increase, but remains roughly the same as that of the third-order
 scheme.  Vlasov--Poisson simulations on $128^3 \times 128^3$ mesh grids
 have been successfully performed on a massively parallel computer.
\end{abstract}

\section{Introduction}
Numerical simulations of collisionless self-gravitating systems are one
of the indispensable tools in the study of galaxies, galaxy clusters, and
the large-scale structure of the universe.  Gravitational $N$-body
simulations have been widely adopted, and significant scientific
achievements have been made for the past four decades. Particles in
$N$-body simulations represent the structure of the distribution
function in the phase space in a statistical manner.  Similarly,
particle-in-cell (PIC) simulations \citep{Birdsall1991} are widely used
for studying the dynamics of an astrophysical plasma such as particle
acceleration in collisionless shock waves \citep{Matsumoto2015} and
magnetic reconnection \citep{Fujimoto2014}.

Particle-based $N$-body simulations have several shortcomings, most
notably the intrinsic shot noise in the physical quantities such as mass
density and velocity field.  The discreteness of the mass distribution
sampled by a finite number of point mass elements (particles) affects
the gravitational dynamics considerably in regions with a small number
of particles, in other words, where the phase space density is low.  For
example, gravitational $N$-body simulations cannot treat collisionless
damping of density fluctuations and the two-stream instability in a
fluid/plasma with a large velocity dispersion because of poor sampling
of particles in the velocity space.  Interestingly, physically
equivalent situations are observed in PIC simulations
\citep{Kates-Harbeck2016}. In PIC simulations, the spacing of grids in
the physical space, on which electric and magnetic fields are calculated
by solving the Maxwell equations, is constrained not to be larger than
the Debye length. Hence, the simulation volume is also restricted for a
given number of grids.

To overcome these shortcomings, a number of alternative methods to the
$N$-body approach have been proposed. \citet{Hernquist1992} and
\citet{Hozumi1997} use the self-consistent field (SCF) method, in which
particles move on trajectories on a smooth gravitational potential
computed by a basis-function expansion of the density field and
gravitational potential. Unlike in conventional $N$-body simulations,
particle motions do not suffer from numerical effects owing to discrete
gravitational potential.  However, a drawback or complication of the SCF
method is that the basis functions need to be suitably chosen for a
given configuration of the self-gravitating systems.

\citet{Abel2012}, \citet{Shandarin2012}, and \citet{Hahn2013} develop a
novel method to simulate cosmic structure formation by following the
motion of dark matter ``sheets'' in the phase space volume, while
retaining the smooth, continuous structure.  The method suppresses
significantly the artificial discreteness effects in $N$-body
simulations such as a spurious filament fragmentation. The method,
however, can only be applied to ``cold'' matter with sufficiently small
velocity dispersions compared with its bulk velocity.
\citet{Vorobyov2006} and \citet{Mitchell2013} solve moment equations of
the collisionless Boltzmann equation in a finite-volume manner by
adopting a certain closure relation. The closure relation needs to be
chosen in a problem-dependent manner, and often the validity of the
adopted relation remains unclear or its applicability is limited.

A straightforward method to simulate a collisionless plasma or a
self-gravitating system is directly solving the collisionless Boltzmann
equation, or the Vlasov equation.  Such an approach was originally
explored by \citet{Janin1971} and \citet{Fujiwara1981} for
one-dimensional self-gravitating systems and by \citet{Nishida1981} for
two-dimensional stellar disks. It is found to be better in following
kinetic phenomena such as collisionless damping and two-stream
instability than conventional $N$-body simulations.

The Vlasov--Poisson and Vlasov--Maxwell equations are solved to simulate
the dynamics of electrostatic and magnetized plasma, respectively.  An
important advantage of this approach is that there is no constraint on
the mesh spacing, unlike in PIC simulations. In principle, direct Vlasov
simulations enable us to simulate the kinematics of an astrophysical
plasma with a wide dynamic range, which cannot be easily achieved with
PIC simulations.

Despite of its versatile nature, the Vlasov simulations have been
applied only to problems in one or two spatial dimensions.  Solving the
Vlasov equation with three spatial dimensions, i.e., in a six-dimensional
phase space, requires an extremely large amount of
memory. \citet{YYU2013} successfully perform Vlasov--Poisson simulations
of self-gravitating systems in six-dimensional phase space,
achieving high accuracies in mass and energy
conservation. Unfortunately, the number of grids is still limited by the
amount of available memory rather than the computational cost. With
currently available computers, it is not realistic to achieve a
significant increase in spatial (and velocity) resolution by simply
increasing the number of mesh grids.  Clearly, developing high-order
schemes is necessary to effectively improve the spatial resolution for a
given number of mesh grids.  In the present paper, we develop and test
new numerical schemes for solving the Vlasov equation based on the fifth- and
seventh-order monotonicity-preserving (MP) schemes \citep{Suresh1997}.
We improve the schemes suitably so that the positivity of the
distribution function is ensured.  We run a suite of test simulations
and compare the results with those obtained by the spatially third-order
positive-flux-conservative (PFC) scheme \citep{Filbet2001} adopted in
our previous work \citep[hereafter YYU13]{YYU2013}.

The rest of the paper is organized as follows. In section 2, we
describe our new numerical schemes to solve the Vlasov equation in
higher-order accuracy in detail. Section 3 is devoted to presenting
several test calculations to show the technical advantage of our
approach over the previous methods. We address the computational costs
of our schemes in section 4. Finally, in section 5, we summarize our
results and present our future prospects.

\section{Vlasov--Poisson Simulation}
We follow the time evolution of the distribution
function $f(\itbold{x}, \itbold{v}, t)$ according to the following
Vlasov (collisionless Boltzmann) equation:
\begin{equation}
 \label{eq:vlasov}
 \pdif{f}{t}+\odif{\itbold{x}}{t}\cdot\pdif{f}{\itbold{x}} +\odif{\itbold{v}}{t} \cdot \pdif{f}{\itbold{v}} = 0,
\end{equation}
where $\itbold{x}=(x_1, x_2, x_3)=(x,y,z)$ and $\itbold{v}=(v_1, v_2,
v_3)=(v_x, v_y, v_z)$ are the Cartesian spatial and velocity
coordinates, respectively. For self-gravitating systems, it gives the
Vlasov--Poisson equation is given as
\begin{equation}
 \label{eq:vlasov-poisson}
 \pdif{f}{t}+\itbold{v}\cdot\pdif{f}{\itbold{x}} -
 \pdif{\phi}{\itbold{x}} \cdot \pdif{f}{\itbold{v}} = 0,
\end{equation}
where $\phi$ is the gravitational potential satisfying the Poisson
equation
\begin{equation}
 \label{eq:poisson}
 \nabla^2\phi = 4\pi G\rho = 4\pi G\int f\, {\rm d}^3\itbold{v},
\end{equation}
where $G$ is the gravitational constant and $\rho$ is the mass density.
Throughout the present paper, the distribution function is normalized
such that its integration over velocity space yields the mass
density. We discretize the distribution function $f(\itbold{x},
\itbold{v}, t)$ in the same manner as in YYU13;
we employ a uniform Cartesian mesh in both the physical configuration
space and in the velocity (momentum) space.

\subsection{Advection Solver}

The Vlasov Equation~(\ref{eq:vlasov-poisson}) can be broken down into
six one-dimensional advection equations along each dimension
of the phase space: three for advection in the physical space
\begin{equation}
 \label{eq:adv_pos}
 \pdif{f}{t} + v_i\pdif{f}{x_i} = 0 \,\,\,\,\,\,\,\,\,\,\,(i=1,2,3),
\end{equation}
and another three in the velocity space
\begin{equation}
 \label{eq:adv_vel}
 \pdif{f}{t} - \pdif{\phi}{x_i}\pdif{f}{v_i} = 0 \,\,\,\,\,\,\,\,\,\,\, (i=1,2,3).
\end{equation}
We solve these six equations sequentially. The
gravitational potential $\phi$ in Equation~(\ref{eq:adv_vel}) is
updated after advection Equation~(\ref{eq:adv_pos}) in the physical
space is solved and the mass density distribution in the physical space
is updated. Clearly, an accurate numerical scheme for a one-dimensional advection
scheme is a crucial ingredient of our Vlasov solver.

Let us consider a numerical scheme of the following advection equation
\begin{equation}
 \label{eq:advection_1d}
 \pdif{f(x,t)}{t} + c \pdif{f(x,t)}{x} = 0,
\end{equation}
on a one-dimensional uniform mesh grid. In a finite-volume manner,
we define an averaged value of $f(x,t)$ over the
$i$th grid centered at $x=x_i$ with the interval of $\Delta x$ at a
time $t=t^n$ as
\begin{equation}
 f^n_i = \frac{1}{\Delta x}\int_{x_{i-1/2}}^{x_{i+1/2}} f(x,
  t^n) \,{\rm d}x,
\end{equation}
where $x_{i\pm 1/2} = x_i \pm \Delta x/2$ are the coordinates of the
boundaries of the $i$th mesh grid.
Without loss of generality, we can
restrict ourselves to the case with a positive advection velocity
$c>0$.

Physical and mathematical considerations impose
the following requirements on numerical solutions of the advection equation:
(i) Monotonicity---since the solution of the advection
Equation~(\ref{eq:advection_1d}) preserves monotonicity, so
should its numerical solution.  (ii) Positivity---the distribution
function must be positive by definition, and thus the numerical
solution should be non-negative. Note that
monotonicity is not sufficient to ensure positivity of the
numerical solution.

\subsubsection{MP Schemes with the Positivity-preserving Limiter}

We first present the spatially fifth-order scheme for the advection
Equation~(\ref{eq:advection_1d}) based on the MP5 scheme of
\citet{Suresh1997}, in which the value of $f(x,t)$ at the interface
between the $(i-1)$th and $i$th mesh grids is reconstructed with
five stencils $f^n_{i-3}$, $f^n_{i-2}$, $f^n_{i-1}$, $f^n_{i}$, and
$f^n_{i+1}$ as
\begin{equation}
 f^{\rm int,5}_{i-1/2} = (2f_{i-3}-13f_{i-2}+47f_{i-1}+27f_i-3f_{i+1})/60.
\end{equation}
In order to ensure monotonicity of the numerical solution,
we impose the constraint
\begin{equation}
 \label{eq:5th_interpolation}
 \bar{f}_{i-1/2} = {\rm MP}(f^{\rm int,5}_{i-1/2}, f_{i-3}, f_{i-2}, f_{i-1}.
  f_{i}, f_{i+1})
\end{equation}
The detailed prescriptions of the constraints
are given in Appendix \ref{appendix:mp5}. The MP5 scheme adopts the
three-stage TVD Runge-Kutta (TVD--RK) time integration as
\begin{eqnarray}
 f^{(1)}_i &=& f^n_i - \nu L_i(f^n) \nonumber\\
 f^{(2)}_i &=& \frac{3}{4}f^n_i + \frac{1}{4}\left(f_i^{(1)}-\nu
  L_i(f^{(1)})\right)\label{eq:RK}\\
 f^{n+1}_i &=& \frac{1}{3}f^n_i + \frac{2}{3}\left(f_i^{(2)}-\nu L_i(f^{(2)})\right)\nonumber,
\end{eqnarray}
where $\nu=c\Delta t/\Delta x$ is the CFL number and the operator
$L_i(f)$ is defined as
\begin{equation}
 L_i(f) = \bar{f}_{i+1/2}-\bar{f}_{i-1/2}.
\end{equation}
Hereafter, this scheme is referred to as RK--MP5.

As already noted above, the MP scheme does not ensure positivity of the
numerical solution.  We thus introduce a positivity-preserving (PP)
limiter based on the method proposed by \citet{Hu2013}. By denoting
$U_i^+ = f_i^n - 2\nu\bar{f}^n_{i+1/2}$ and $U_i^-=
f^n_i+2\nu\bar{f}^n_{i-1/2}$, the single-stage Eulerian time integration
can be written as
\begin{eqnarray}
 f^{(1)}_i & = & f^n_i - \nu(\bar{f}^n_{i+1/2}-\bar{f}^n_{i-1/2})\\
 & = &
  \frac{1}{2}\left(f^n_i-2\nu\bar{f}^n_{i+1/2}\right)+\frac{1}{2}\left(f^n_i+2\nu\bar{f}^n_{i-1/2}\right)\\
 & = & \frac{1}{2} U_i^+ + \frac{1}{2} U_i^-.
\end{eqnarray}
Here, we consider a linear combination of the interface value obtained
with the MP scheme and that with the spatially first-order upwind scheme as
\begin{equation}
 \label{eq:combined_flux}
 \hat{f}^n_{i+1/2} = \theta_{i+1/2}\bar{f}^n_{i+1/2} +
  (1-\theta_{i+1/2})f^{{\rm UP},n}_{i+1/2},
\end{equation}
where $f^{{\rm UP},n}_{i+1/2}$ is expressed as
\begin{equation}
 f^{{\rm UP},n}_{i+1/2} = \frac{1}{2c}\left[f^n_i(c+|c|)+f^n_{i+1}(c-|c|)\right] = f_i^n,
\end{equation}
and $\theta_{i+1/2}$ is a parameter with a condition of
$0\le\theta_{i+1/2}\le 1$. Note that the spatially first-order upwind
scheme preserves positivity of the numerical solution if the CFL
condition $\nu\le 1$ is satisfied. Then, the time integration of the
combined interface values~(\ref{eq:combined_flux}) can be written as
\begin{eqnarray}
 f^{(1)}_i&=& f_i^n- \nu(\hat{f}^n_{i+1/2}-\hat{f}^n_{i-1/2})\\
 &=&
  \frac{1}{2}\left(f^n_i-2\nu\hat{f}^n_{i+1/2}\right)+\frac{1}{2}\left(f^n_i+2\nu\hat{f}^n_{i-1/2}\right)\\
 &=& \frac{1}{2}\left\{\theta_{i+1/2}U^+_i+(1-\theta_{i+1/2})U^{\rm
		 UP,+}_i\right\}+\frac{1}{2}\left\{\theta_{i-1/2}U^-_i +
		 (1-\theta_{i-1/2})U^{\rm UP,-}_i\right\}\label{eq:combined},
\end{eqnarray}
where $U_i^{\rm UP,+} = f_i^n-2\nu f^{{\rm UP},n}_{i+1/2}$ and $U_i^{\rm
UP,-}=f_i^n + 2\nu f^{{\rm UP},n}_{i-1/2}$. Therefore, if both of the
two terms on the right-hand side of Equation~(\ref{eq:combined}) are positive,
$f^{(1)}_i$ is also positive. A simple calculation shows that
it can be realized by setting the parameter $\theta_{i+1/2}$ as
\begin{equation}
 \theta_{i+1/2} = \min(\theta_{i+1/2}^+, \theta_{i+1/2}^-),
\end{equation}
where
\begin{equation}
 \theta_{i+1/2}^+ = \left\{
		     \begin{array}{ll}
		      \displaystyle \frac{U_i^{\rm UP,+}}{U_i^{\rm UP,+}-U_i^+} &
		       \,\,\,\,\mbox{if $U_i^+<0$}\\
		      1 & \,\,\,\,\mbox{otherwise}
		     \end{array}
		     \right.
\end{equation}
and
\begin{equation}
 \theta_{i+1/2}^- = \left\{
		     \begin{array}{ll}
		      \displaystyle \frac{U_{i+1}^{\rm UP,-}}{U_{i+1}^{\rm UP,-}-U_{i+1}^-} &
		       \,\,\,\,\mbox{if $U_{i+1}^-<0$}\\
		      1 & \,\,\,\,\mbox{otherwise.}
		     \end{array}
		     \right.
\end{equation}
In addition, the condition of $\theta_{i+1/2} \ge 0$ can be met by
setting $U_i^{\rm UP,+}\ge 0$ and $U_{i+1}^{\rm UP,-}\ge 0$ as a
sufficient condition, which leads to a constraint on the CFL parameter
as $\nu\le1/2$.
Let us denote the above procedures to evaluate the positive
interface value~(\ref{eq:combined_flux}) as
\begin{equation}
 \hat{f}^n_{i+1/2} = {\rm PP}(\bar{f}^n_{i+1/2}, f^n_i, f^n_{i+1}).
\end{equation}
It can be shown that positivity of the numerical solution is
preserved even with the three-stage TVD--RK time integration by setting the
CFL parameter to $\nu < 1/2$.  In the following, we call this scheme
the RK--MPP5 scheme (MPP for ``monotonicity and positivity
preserving'').

It is tedious but straightforward to construct
the spatially seventh-order schemes, RK--MP7 and RK--MPP7,
by simply replacing $f_{i-1/2}^{\rm int,5}$ in
Equation~(\ref{eq:5th_interpolation}) with the seventh-order interpolation
\begin{equation}
 \label{eq:7th_interpolation}
 f_{i-1/2}^{\rm int,7} = (-3f_{i-4}+25f_{i-3}-101f_{i-2}+319f_{i-1}+214f_i-38f_{i+1}+4f_{i+2})/420.
\end{equation}

\subsubsection{Semi-Lagrangian Time Integration}

The MP schemes described above adopt the three-stage TVD--RK time
integration~(\ref{eq:RK}), which needs explicit Euler integration three times
per single time step. It can be a source of significant
computational cost. To overcome this
drawback, we devise an alternative method based on the
semi-Lagrangian (SL) time integration.

\begin{figure}[htbp]
 \centering
 \includegraphics[width=10cm]{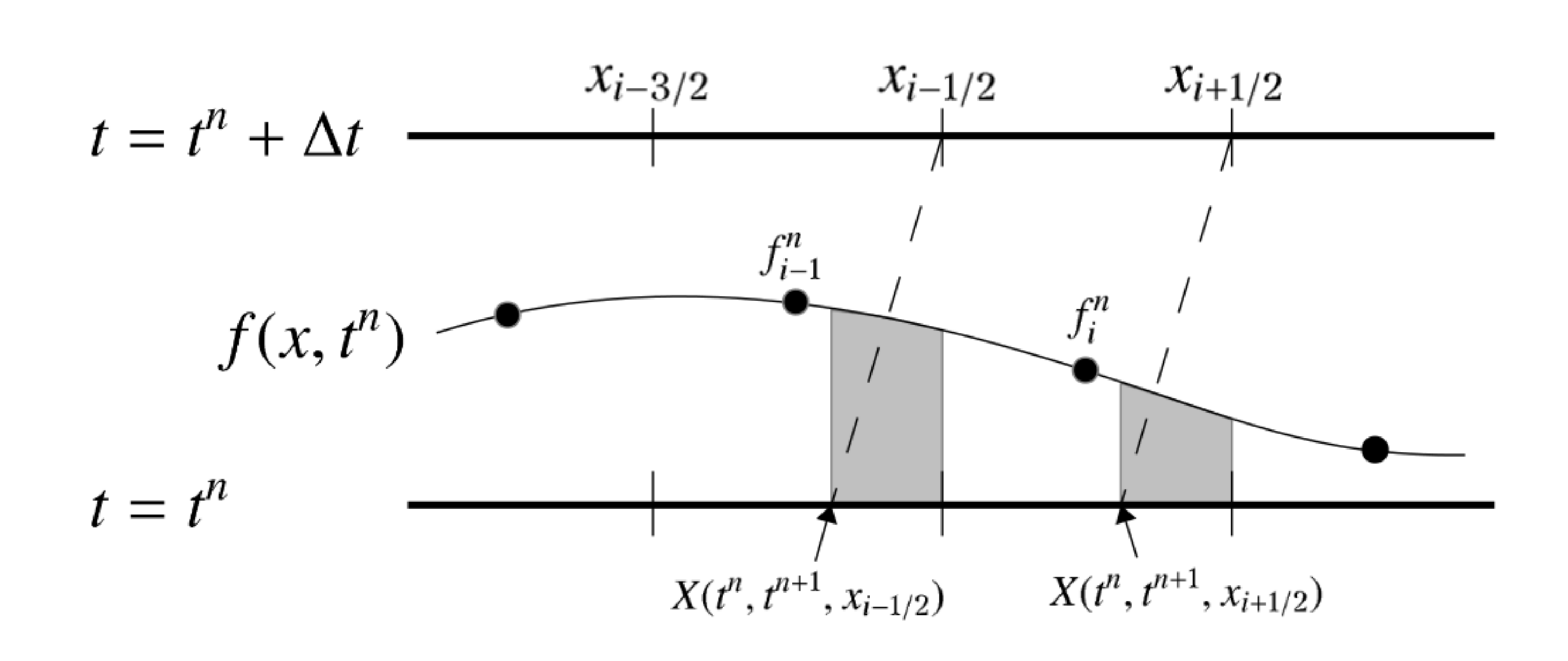}
 \figcaption{Schematic description of the semi-Lagrangian scheme. Dashed
 lines starting from $x_{i\pm 1/2}$ are the characteristic curves. The
 areas of the left and right shaded regions correspond to $\Phi_{i-1/2}$
 and $\Phi_{i+1/2}$, respectively.}
 \label{fig:sl}
\end{figure}

The averaged value of $f(x,t)$ over the $i$th mesh grid $f^{n+1}_i$ at
a time of $t^{n+1} = t^n + \Delta t$ is written as
\begin{equation}
 \label{eq:csl_evolution}
 f^{n+1}_i = \frac{1}{\Delta x}\int_{x_{i-1/2}}^{x_{i+1/2}}
  f(x, t^{n+1})\,{\rm d}x = \frac{1}{\Delta x}\int_{X(t^n, t^{n+1},
  x_{i-1/2})}^{X(t^n, t^{n+1}, x_{i+1/2})} f(x, t^n) \,{\rm d}x,
\end{equation}
where $X(t_1, t_2, x)$ is the position of the characteristic curve at a
time of $t_1$ originating from the phase space coordinate $(x, t_2)$,
and for the advection Equation~(\ref{eq:advection_1d}),
$X(t_1, t_2, x)=x-c(t_2-t_1)$; see Figure~\ref{fig:sl}. Denoting
\begin{equation}
 \Phi^n_{i+1/2} = \frac{1}{c\Delta t}\int_{X(t^n, t^{n+1}, x_{i+1/2})}^{x_{i+1/2}} f(x, t^n)
  \,{\rm d}x
\end{equation}
and
\begin{equation}
 \Phi^n_{i-1/2} = \frac{1}{c\Delta t}\int^{x_{i-1/2}}_{X(t^n, t^{n+1}, x_{i-1/2})} f(x, t^n)
  \,{\rm d}x,
\end{equation}
we can rewrite the Equation~(\ref{eq:csl_evolution}) in terms of
numerical flux as
\begin{equation}
 f^{n+1}_i = f^n_i - \nu(\Phi^n_{i+1/2} - \Phi^n_{i-1/2}) .
\end{equation}
By interpolating the discrete values of $f^n_j$, we can compute
the numerical fluxes $\Phi_{i+1/2}$ and $\Phi_{i-1/2}$. Actually, the
PFC scheme also adopts essentially the same approach.
It adopts the third-order reconstruction to interpolate $f^n_j$ with
a slope limiter and suppresses numerical oscillation while preserving
positivity of the numerical solutions. For more detailed description
of the PFC scheme, see \citet{Filbet2001} and the Appendix of
YYU13.

Our new schemes are based on the conservative semi-Lagrangian (CSL) scheme
\citep{Qiu2010, Qiu2011}, which is originally coupled with a finite-difference
scheme but is also applicable to the finite-volume schemes. In the
fifth-order CSL scheme, $\Phi^n_{i-1/2}$ can be expressed with five
stencil values as
\begin{equation}
 \Phi^n_{i-1/2} = \sum_{j=0}^{4} C_j\zeta^j
\end{equation}
where
\begin{equation}
 C_0 = \frac{f^n_{i-3}}{30}-\frac{13}{60}f^n_{i-2}+\frac{47}{60}f^n_{i-1}+\frac{9}{20}f^n_{i}-\frac{f^n_{i+1}}{20}
\end{equation}
\begin{equation}
 C_1 = -\frac{f^n_{i-2}}{24} + \frac{5}{8}f^n_{i-1} - \frac{5}{8}f^n_{i} + \frac{f^n_{i+1}}{24}
\end{equation}
\begin{equation}
 C_2 = -\frac{f^n_{i-3}}{24} + \frac{f^n_{i-2}}{4} - \frac{f^n_{i-1}}{3} +
  \frac{f^n_{i}}{12} + \frac{f^n_{i+1}}{24}
\end{equation}
\begin{equation}
 C_3 = \frac{f^n_{i-2}}{24} - \frac{f^n_{i-1}}{8} + \frac{f^n_i}{8} - \frac{f^n_{i+1}}{24}
\end{equation}
\begin{equation}
 C_4 = \frac{f^n_{i-3}}{120} - \frac{f^n_{i-2}}{30} + \frac{f^n_{i-1}}{20} -
  \frac{f^n_i}{30} + \frac{f^n_{i+1}}{120}
\end{equation}
and
\begin{equation}
 \zeta = \frac{c\Delta t}{\Delta x}.
\end{equation}
The seventh-order CSL scheme is given by
\begin{equation}
 \Phi^n_{i-1/2} = \sum_{j=0}^6 D_j\zeta^j,
\end{equation}
where
\begin{equation}
 D_0 = -\frac{f^n_{i-4}}{140} + \frac{5}{84}f^n_{i-3} - \frac{101}{420}f^n_{i-2} +
  \frac{319}{420}f^n_{i-1} + \frac{107}{210}f^n_{i} -
  \frac{19}{210}f^n_{i+1} + \frac{f^n_{i+2}}{105}
\end{equation}
\begin{equation}
 D_1 = \frac{f^n_{i-3}}{180} - \frac{5}{72}f^n_{i-2} +
  \frac{49}{72}f^n_{i-1} - \frac{49}{72}f^n_{i} +
  \frac{5}{72}f^n_{i+1} - \frac{f^n_{i+2}}{180}
\end{equation}
\begin{equation}
 D_2 = \frac{7}{720}f^n_{i-4} - \frac{19}{240}f^n_{i-3} + \frac{7}{24}f^n_{i-2} -
  \frac{23}{72}f^n_{i-1} + \frac{f^n_{i}}{48} +
  \frac{7}{80}f^n_{i+1} - \frac{f^n_{i+2}}{90}
\end{equation}
\begin{equation}
 D_3 = - \frac{f^n_{i-3}}{144} + \frac{11}{144}f^n_{i-2} -
  \frac{7}{36}f^n_{i-1} + \frac{7}{36}f^n_{i} -
  \frac{11}{144}f^n_{i+1} + \frac{f^n_{i+2}}{144}
\end{equation}
\begin{equation}
 D_4 = -\frac{f^n_{i-4}}{360} + \frac{f^n_{i-3}}{48} - \frac{13}{240}f^n_{i-2} +
  \frac{23}{360}f^n_{i-1} - \frac{f^n_{i}}{30} +
  \frac{f^n_{i+1}}{240} + \frac{f^n_{i+2}}{720}
\end{equation}
\begin{equation}
 D_5 = \frac{f^n_{i-3}}{720} - \frac{f^n_{i-2}}{144} +
  \frac{f^n_{i-1}}{72} - \frac{f^n_{i}}{72} +
  \frac{f^n_{i+1}}{144} - \frac{f^n_{i+2}}{720}
\end{equation}
and
\begin{equation}
 D_6 = \frac{f^n_{i-4}}{5040} - \frac{f^n_{i-3}}{840} + \frac{f^n_{i-2}}{336} -
  \frac{f^n_{i-1}}{252} + \frac{f^n_{i}}{336} -
  \frac{f^n_{i+1}}{840} + \frac{f^n_{i+2}}{5040}.
\end{equation}
However, the numerical solution obtained with the CSL scheme does not
preserve the monotonicity. Thus, we apply the MP constraint to
$\Phi^n_{i-1/2}$ and $\Phi^n_{i+1/2}$ as
\begin{equation}
 \bar{\Phi}^n_{i-1/2} = {\rm MP}(\Phi^n_{i-1/2}, f^n_{i-3}, f^n_{i-2},
  f^n_{i-1}, f^n_{i}, f^n_{i+1})
\end{equation}
and
\begin{equation}
 \bar{\Phi}^n_{i+1/2} = {\rm MP}(\Phi^n_{i+1/2}, f^n_{i-2}, f^n_{i-1},
  f^n_{i}, f^n_{i+1}, f^n_{i+2}).
\end{equation}
The numerical solution with the monotonicity preservation at
$t=t^{n+1}$ is then computed as
\begin{equation}
 f^{n+1}_i = f^n_i - \nu(\bar{\Phi}^n_{i+1/2}-\bar{\Phi}^n_{i-1/2}),
\end{equation}
which is referred to as the SL--MP5 and SL--MP7 schemes (SL for
``semi-Lagrangian'') for the fifth- and seventh-order accuracies,
respectively.

The PP limiter described in the
previous subsection can be applied as
\begin{equation}
 \hat{\Phi}^n_{i-1/2} = {\rm PP}(\bar{\Phi}^n_{i-1/2}, f^n_{i-1}, f^n_i)
\end{equation}
and
\begin{equation}
 \hat{\Phi}^n_{i+1/2} = {\rm PP}(\bar{\Phi}^n_{i+1/2}, f^n_i, f^n_{i+1}).
\end{equation}
Then, the numerical solution at $t^{n+1}$ is given by
\begin{equation}
 \label{eq:SL}
 f^{n+1}_i = f^n_i - \nu(\hat{\Phi}^n_{i+1/2} - \hat{\Phi}^n_{i-1/2}).
\end{equation}
We refer to these schemes as the SL--MPP5 and SL--MPP7 schemes hereafter.

\subsection{Poisson Solver}

The Poisson Equation~(\ref{eq:poisson}) is numerically solved with the
convolution method \citep{Hockney1981} using the fast Fourier transform
(FFT). Under periodic boundary conditions, the discrete Fourier transform
(DFT) of the gravitational potential $\phi(\itbold{k})$ can be written
as
\begin{equation}
 \hat{\phi}({\itbold{k}}) = G(\itbold{k})\hat{\rho}(\itbold{k})
\end{equation}
where $\hat{\rho}(\itbold{k})$ is the DFT of the density field,
$\itbold{k}=(k_x, k_y, k_z)$ is the discrete wavenumber vector, and
$G(\itbold{k})$ is the Fourier-transformed Green function of the
discretized Poisson equation. The second-order discretization of the
Green function, $G(\itbold{k})$, is given by
\begin{equation}
 G(\itbold{k})= -\frac{\pi G \Delta^2}{\displaystyle \sum_{i=x,y,z}\sin^2\left(\frac{k_i\Delta}{2}\right)}
\end{equation}
with $\Delta x = \Delta y = \Delta z = \Delta$. Since we adopt
advection solvers with spatially fifth- and seventh-order
accuracy, it is reasonable to adopt a Poisson solver with the same or
similar order of
accuracy, which can be simply constructed by high-order
discretization of the Poisson equation. As shown in Appendix
\ref{appendix:poisson_solver}, the DFTs of the Green functions with
spatially fourth- and sixth-order accuracy are given by
\begin{equation}
 G(\itbold{k}) = \frac{12\pi G \Delta^2}{\displaystyle\sum_{i=x,y,z}\left[\sin^2(k_i\Delta)-16\sin^2\left(\frac{k_i\Delta}{2}\right)\right]}
\end{equation}
and
\begin{equation}
 G(\itbold{k}) = -\frac{180\pi G
  \Delta^2}{\displaystyle\sum_{i=x,y,z}\left[2\sin^2\left(\frac{3k_i\Delta}{2}\right)-27\sin^2(k_i\Delta)+270\sin^2\left(\frac{k_i\Delta}{2}\right)\right]},
\end{equation}
respectively. The gravitational potential $\phi(\itbold{x})$ is computed
with the inverse DFT of $\hat{\phi}(\itbold{k})$.

For isolated boundary conditions, the doubling-up method
\citep{Hockney1981} is adopted, where the number of mesh grids is
doubled for each spatial dimension and the mass density in the extended
grid points is zeroed out. The Green function is set up in the real
space as
\begin{equation}
 G(x, y, z) = \frac{G}{(x^2 + y^2 + z^2)^{1/2}}
\end{equation}
and is Fourier-transformed to obtain $\hat{G}(\itbold{k})$. Then, the
gravitational potential can be calculated in the same manner as in the
the periodic boundary condition.

After obtaining the gravitational potential, we compute gravitational
forces at each of the spatial grids with the finite-difference
approximation (FDA). Again we adopt FDA schemes with the same order of
accuracy as that of the gravitational potential calculation.
The two-, four- and six-point FDAs are described in Appendix
\ref{appendix:poisson_solver}.  In what follows, we adopt the
sixth-order calculation of the gravitational forces and potentials
irrespective of adopted advection schemes because we are primarily
interested in the accuracy of our advection equation solver.

\subsection{Parallelization and Domain Decomposition}
We parallelize our Vlasov--Poisson solvers by decomposing the six-dimensional phase
space into distributed memory spaces. We decompose the computational
domain in the phase space in the same manner as YYU13. The
physical space is divided into rectangular subdomains along $x$-, $y$-,
and $z$-directions, while the velocity space is not decomposed;
each spatial point has an entire set of velocity mesh grids.
This decomposition is
convenient when computing the velocity moments of the distribution function
such as the mass density, mean velocities, and velocity dispersions.

In solving the advection equations in the physical space
(\ref{eq:adv_pos}) across the decomposed subdomains, the values of the
distribution function adjacent to the boundaries
of the subdomains are transferred using the Message Passing Interface
(MPI). The number of the adjacent mesh grids to be transferred is two for
the spatially third-order PFC scheme and three and four for spatially
fifth- and seventh-order MP schemes, respectively.

\subsection{Time Integration}
\label{sec:time_integration}
We advance the distribution function from $t=t^n$ to $t^{n+1}$ using
the directional splitting method on the order of
\begin{eqnarray}
 \label{eq:splitting}
  f(\itbold{x},\itbold{v}, t^{n+1})& =& T_{v_z}(\Delta t/2) T_{v_y}(\Delta t/2) T_{v_x}(\Delta t/2) \nonumber \\
  &&\times T_{x}(\Delta t)T_{y}(\Delta t)T_{z}(\Delta t) \nonumber\\
  &&\times T_{v_z}(\Delta t/2) T_{v_y}(\Delta t/2) T_{v_x}(\Delta t/2) f(\itbold{x},\itbold{v}, t^n),
\end{eqnarray}
where $T_{\ell} (\Delta t)$ denotes an operator for advection along the
$\ell$-direction with time step $\Delta t$. It should be noted that the
Poisson equation is solved after the advection operations in the
physical space 
($T_{x} T_{y} T_{z}$) to update the gravitational potential and forces
for the updated density field. We adopt the directional splitting
method rather than the unsplit method because the former is the
combination of one-dimensional advection schemes and hence rigorously
assures the monotonicity and the positivity even in the
multidimensional problems and also because the latter requires a larger
amount of memory space to temporally store the numerical fluxes or the
updated distribution function. Furthermore, the unsplit method requires
a relatively smaller time step than the directional splitting method to
achieve the same level of numerical accuracy. See
section~\ref{subsec:2D_advection} for the comparison of the dimensional
splitting and unsplit methods in the two-dimensional advection
problems. The temporal accuracy of the directional splitting method is
in the second order at best. However, we find that the errors in the
numerical solutions presented in the following numerical test are
mostly dominated by ones that originate from the spatial discretization, and
the spatial resolution is more important than the temporal accuracy in
terms of the quality of the numerical solution.

As for time step width, we have several constraints on the CFL
parameter. The MP schemes are subject to the constraint with $\nu\le
1/5$ if the parameter $\alpha$ in the MP scheme is set to $\alpha=4$
(see Appendix \ref{appendix:mp5}) and the PP limiter requires $\nu\le
1/2$. As shown in Appendix~\ref{appendix:cfl}, all of our new schemes
work well for $3<\alpha$ and $\nu<0.4$. In practice, the condition
$\nu=1/(1+\alpha)$ is desirable, and thus we set $\alpha=4$ and
$\nu=0.2$ throughout the present paper unless otherwise stated.
\footnote{The PFC scheme is free from any explicit constraints on time
step thanks to its semi-Lagrangian nature in solving the one-dimensional
advection equation. Nonetheless, it is better to adopt a certain
constraint in domain-decomposed (and hence parallelized) Vlasov
simulations in multidimensional phase space. With large time steps, the
trajectories of characteristic lines originating from the boundaries of
decomposed domains reach further from the boundaries, and the number of
grids to be passed to the adjacent domains becomes also larger, resulting
in the increase of the amount of data exchange between parallel
processes.}

We set the time step for integrating the Vlasov equation as
\begin{equation}
  \Delta t = \min(\Delta t_{\rm p}, \Delta t_{\rm v}),
\end{equation}
where $\Delta t_{\rm p}$ and $\Delta t_{\rm v}$ are the time steps for
advection equations in physical and velocity space,
respectively. $\Delta t_{\rm p}$ is given by
\begin{equation}
  \Delta t_{\rm p} = \nu \min\left(\frac{\Delta x}{V_x^{\rm max}},\frac{\Delta y}{V_y^{\rm max}},\frac{\Delta z}{V_z^{\rm max}}\right),
\end{equation}
where $\Delta \ell$ denotes the spacing of mesh grids along the
$\ell$-direction and $V_{\ell}^{\rm max}$ is the maximum extent of the
velocity space along the $v_{\ell}$-direction. Similarly, we compute $\Delta
t_{\rm v}$ as
\begin{equation}
  \Delta t_{\rm v} = \nu \min_i\left(\frac{\Delta v_x}{|a_{x,i}|},
                                   \frac{\Delta v_y}{|a_{y,i}|},
                                   \frac{\Delta v_z}{|a_{z,i}|}\right),
\end{equation}
where $a_{\ell,i}$ is the gravitational acceleration along the
$\ell$-direction at the $i$th mesh grids in the physical space and
the minimization is taken over all the spatial mesh grids.

\section{Test Calculations}

In this section, we present the results of several test simulations with
our new schemes.  The test suite includes simulations of one- and
two-dimensional advection and Vlasov--Poisson simulations in two- and
six-dimensional phase space.

\subsection{Test-1: Advection Equation}

As the first of the test suite of our new schemes, we perform
numerical simulations of the advection equation. Here, we solve the one-
and two-dimensional advection equations. In the one-dimensional tests,
we examine the fundamental properties of our new schemes such as
monotonicity, positivity, and accuracy. In the two-dimensional tests, we
present the applicability of our one-dimensional schemes to
multidimensional problems.

The first test run we perform is the one-dimensional advection problem.
We solve
\begin{equation}
 \label{eq:1d_adv}
 \pdif{f(x,t)}{t} + \pdif{f(x,t)}{x} = 0
\end{equation}
with our new schemes. Since the accuracy of the numerical solutions depends
on the shape of $f(x,t)$, we consider three initial conditions and examine
critically the numerical solutions.

\subsubsection{Test-1a: Rectangular-shaped Function}
\begin{figure}[htbp]
 \begin{center}
  \includegraphics[width=15cm]{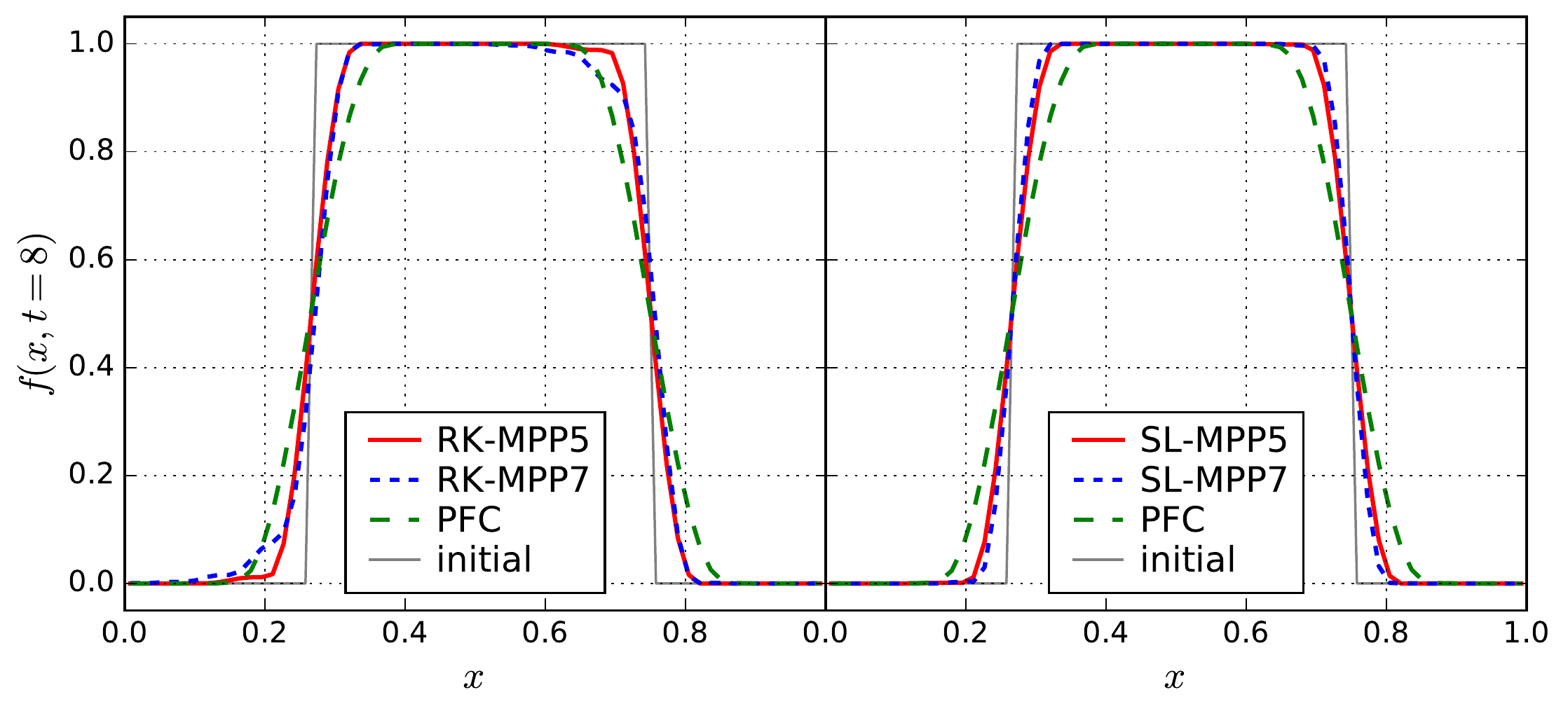}
  \figcaption{Test-1a (rectangular-shaped function): profiles of the
  initially rectangular-shaped function at $t=8$ computed with PFC,
  RK--MPP5, RK--MPP7, SL--MPP5, and SL--MPP7. $N_{\rm m}$ is set to 64.
  \label{fig:1d_adv_rect}}
 \end{center}
\end{figure}

\begin{figure}[htbp]
 \centering \includegraphics[width=10cm]{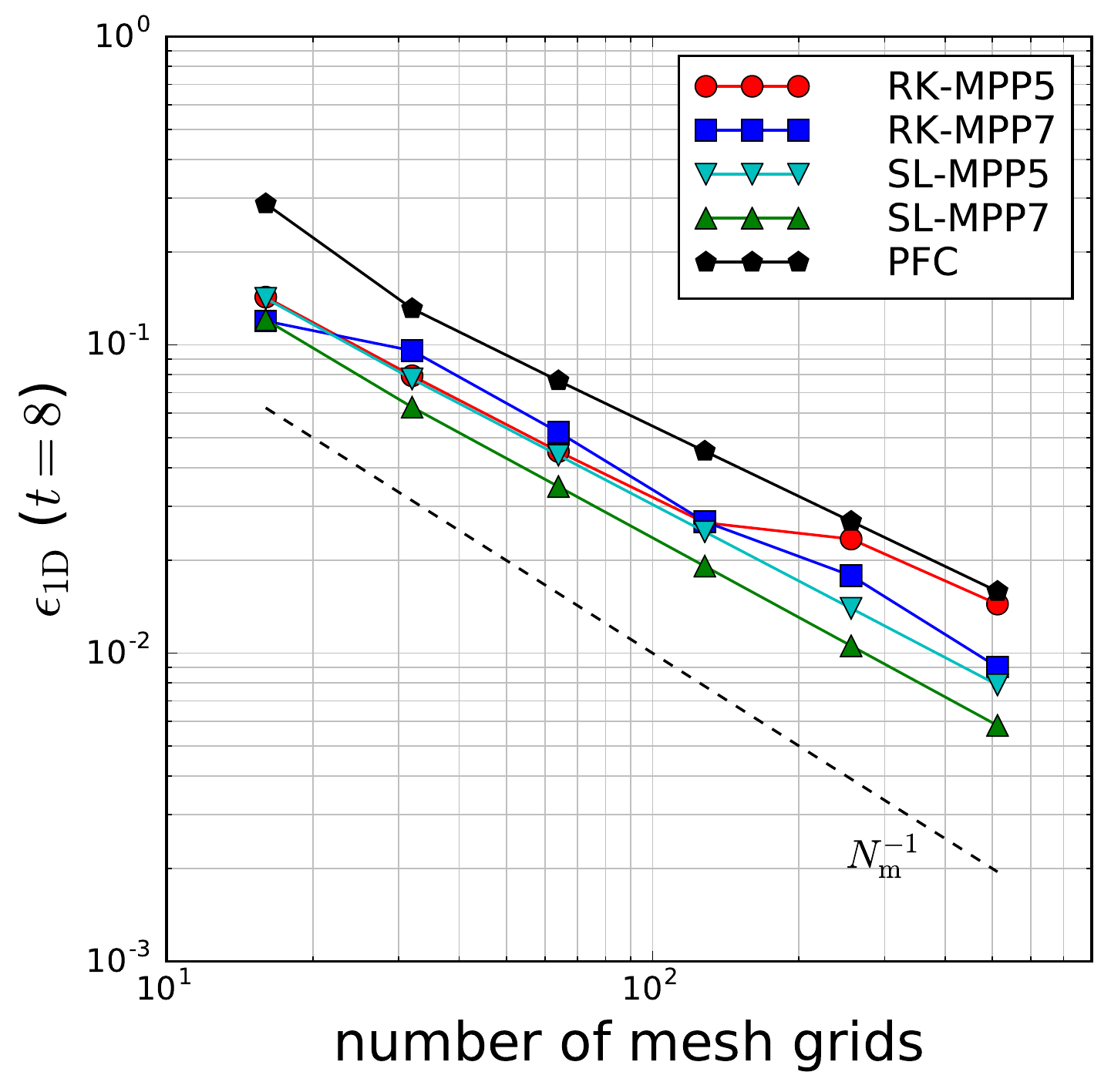}
 \figcaption{Test-1a (rectangular-shaped function): relative errors of
 the numerical solutions computed with PFC, RK--MPP5, RK--MPP7, SL--MPP5,
 and SL--MPP7. A thin dashed line indicates the error scaling
 proportional to $N_{\rm m}^{-1}$. \label{fig:1d_adv_rect_error}}
\end{figure}

We configure the rectangular-shaped initial condition given by
\begin{equation}
 f(x,t=0) = \left\{\begin{array}{lll}
	     1 && 0.25\le x \le 0.75\\
	      0 && {\rm otherwise}
		 \end{array}\right.
\end{equation}
in the simulation domain over $0\le x \le 1$ and integrate equation
(\ref{eq:1d_adv}) using PFC, RK--MPP5, RK--MPP7, SL--MPP5, and SL--MPP7
under the periodic boundary conditions. Figure~\ref{fig:1d_adv_rect}
shows the numerical solutions at $t=8$ with the number of mesh grids
$N_{\rm m} = 64$.  The numerical solutions with RK--MPP5, SL--MPP5, and
SL--MPP7 are less diffusive compared with that of PFC. Note that
RK--MPP7 yields undesirable asymmetric features around $x\simeq 0.2$ and
$0.7$. Actually these features are persistent irrespective of $N_{\rm
m}$ and $\nu$, and are likely caused by the lack of accuracy in the time
integration. In order to avoid these features, we would need the fourth-
or higher-order TVD--RK integration schemes, which are computationally
quite expensive (see \S~\ref{sec:computational_cost}). Thus, we exclude
RK--MP7 and RK--MPP7 in the rest of our test simulations. Unlike
RK--MPP7, SL--MPP7 does not exhibit such features, indicating that the
SL time integration method is able to be combined with spatially
seventh- or higher-order schemes. We confirm that the spatially
ninth-order CSL scheme combined with the SL method works well at least
in one-dimensional advection problems.
Figure~\ref{fig:1d_adv_rect_error} shows the relative error of the
numerical solutions to the analytic solution defined as
\begin{equation}
 \label{eq:L1_error}
 \epsilon_{1{\rm D}} =  \frac{\displaystyle\sum_{i=1}^{N_{\rm m}} |f_i -
  f(x_i,t)| }{\displaystyle\sum_{i=1}^{N_{\rm m}} |f(x_i,t)|}.
\end{equation}
Since there are discontinuities in the initial condition, the numerical
fluxes computed with PFC and our new schemes are effectively of the
first order, avoiding numerical oscillations. This is why the relative
errors scale with the first-order power law.

\subsubsection{Test-1b: Sinusoidal Wave}
\begin{figure}[htbp]
 \centering
 \includegraphics[width=15cm]{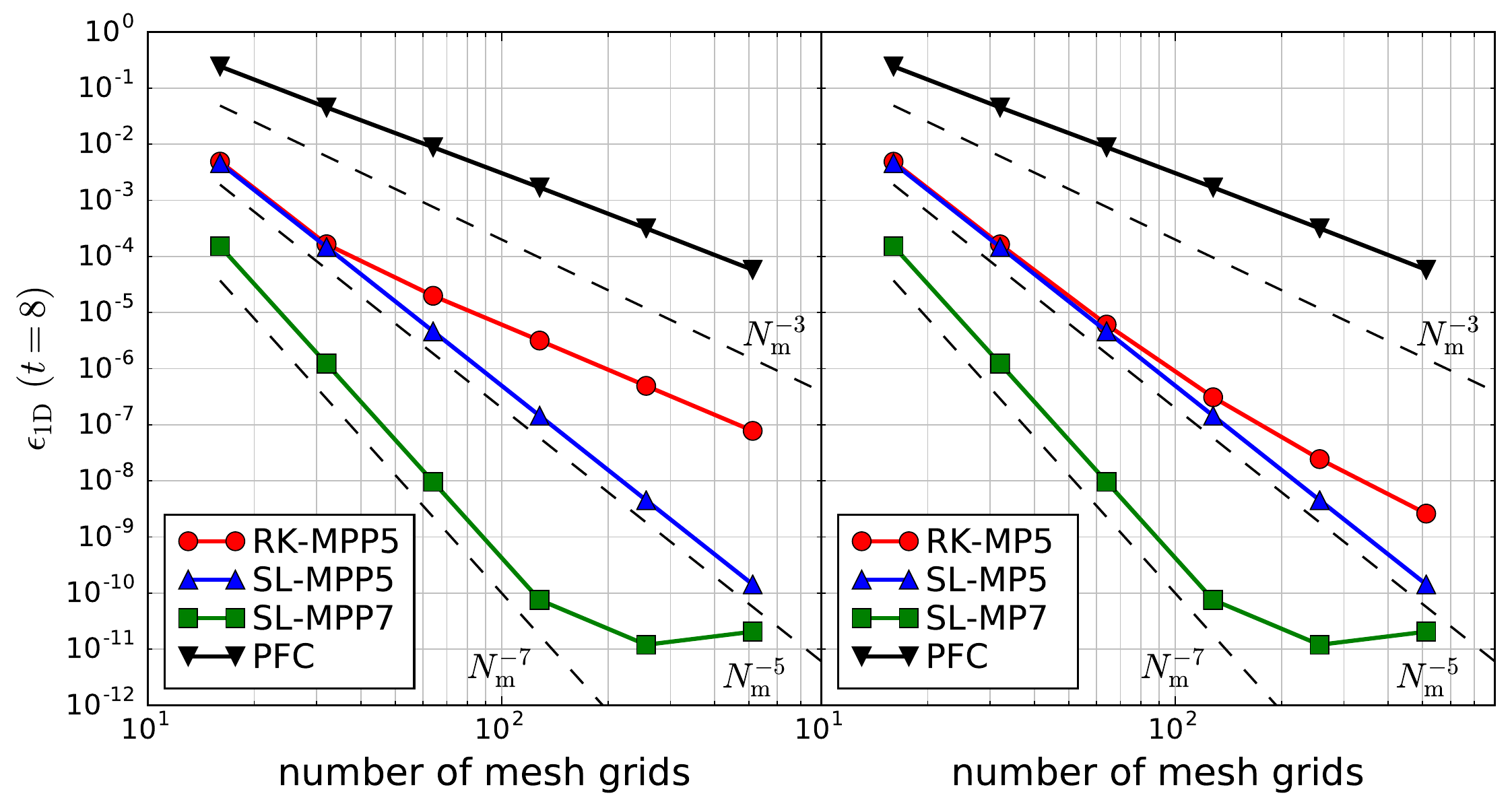}
 \figcaption{Test-1b (sinusoidal wave): relative errors of the advection
 of sinusoidal waves computed with schemes with and without the
 PP limiter in the left and right panels,
 respectively. Thin dashed lines in each panels indicate the scalings of
 the numerical errors in the third-, fifth-, and seventh-order accuracies
 from top to bottom. \label{fig:1d_adv_sin_error}}
\end{figure}

As a good example of advection of a smooth function, we consider
the initial condition given by
\begin{equation}
 f(x,t=0) = 1+\sin(2\pi x),
\end{equation}
in the periodic domain of $0\le x\le 1$.
Figure~\ref{fig:1d_adv_sin_error} shows the relative errors of numerical
solutions obtained with RK--MPP5, SL--MPP5, and SL--MPP7 in the left
panel and with RK--MP5, SL--MP5, and SL--MP7 without the PP
limiter in the right panel as a function of $N_{\rm m}$.
For comparison, the result obtained with PFC is also presented in both
the panels. The numerical solutions obtained with RK--MPP5 have nearly
third-order accuracy for $N_{\rm m}\gtrsim 10^2$, whereas those with
RK--MP5 have fifth-order accuracy. This is because the PP limiter
operates at the minima of numerical solutions at each stage of the
TVD--RK time integration and degrades overall accuracy.  Contrastingly,
the accuracies of the SL schemes are not affected by the PP limiter,
i.e., the results of SL--MPP5 and SL--MPP7 show fifth- and seventh-order
scaling.

\subsubsection{Test-1c: Quartic Sine Function}

\begin{figure}[htbp]
 \centering \includegraphics[width=15cm]{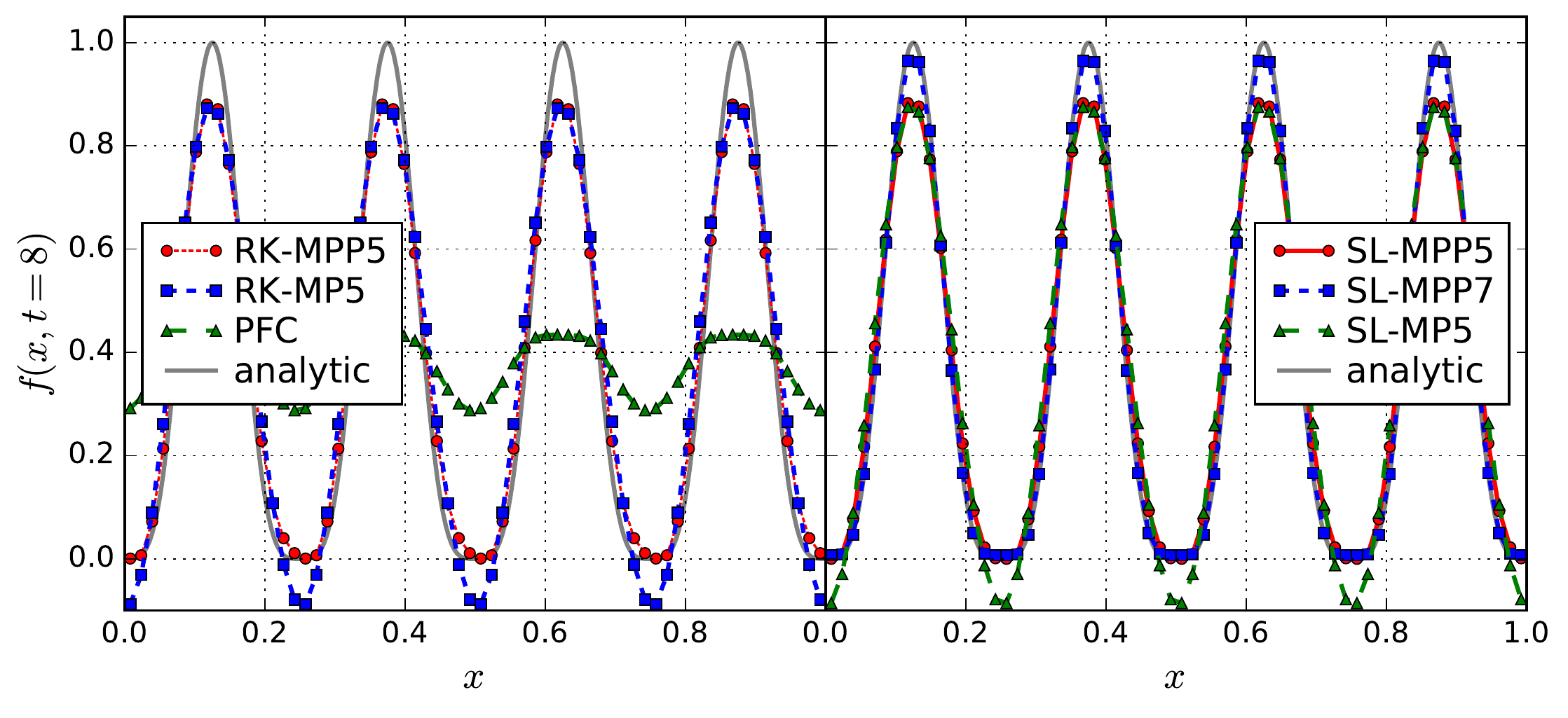}

 \figcaption{Test-1c (quartic sine function): profiles of the numerical
 solutions at $t=8$ computed with the RK schemes (left panel) and the
 SL schemes (right panel). For comparison, we present the result
 obtained with PFC in the left panel. $N_{\rm m}$n is set to
 64. \label{fig:1d_adv_sin4}}
\end{figure}

\begin{figure}[htbp]
 \centering \includegraphics[width=10cm]{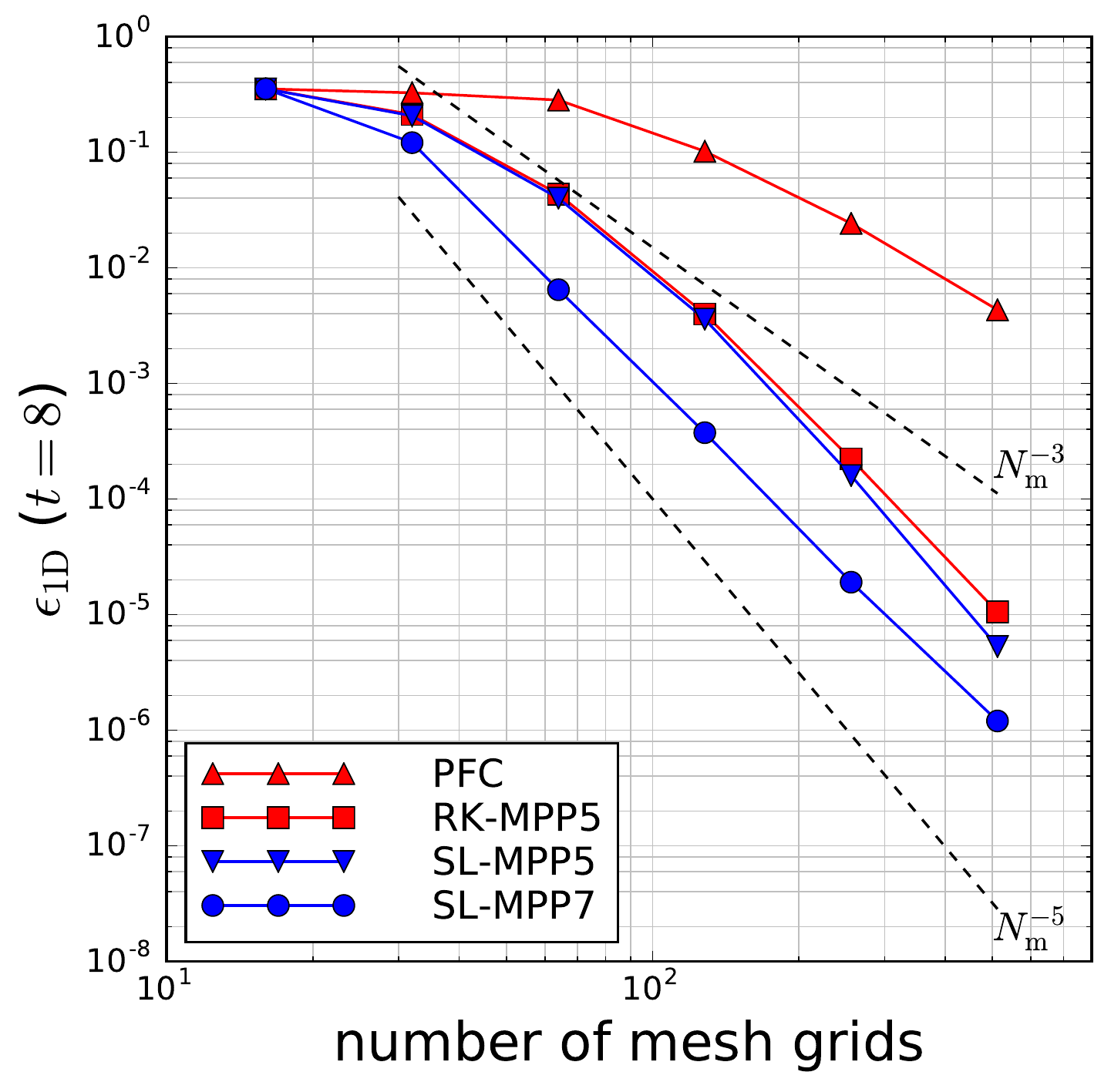}

 \figcaption{Test-1c (quartic sine function): relative errors of PFC,
 RK--MPP5, SL--MPP5, and SL--MP7. Thin dashed lines show the
 error scaling of third- and fifth-order accuracies from top to bottom.
 \label{fig:1d_adv_sin4_error}}
\end{figure}

The final example is the initial condition given by
\begin{equation}
 f(x,t=0) = \sin^4(4\pi x).
\end{equation}
This case serves as a good example to examine how well the PP limiter
works. Figure~\ref{fig:1d_adv_sin4} shows the numerical solutions with
$N_{\rm m}=64$ at $t=8$ computed with the PFC, RK--MP5, and RK--MPP5
(left panel) and those obtained with SL--MP5, SL--MPP5, and SL--MPP7
(right panel).  The PFC solution is significantly smeared near the
maxima and minima, although it preserves positivity. The numerical
solutions obtained with RK--MP5 and SL--MP5 without the PP limiter
exhibit noticeable negative values around the minima. On the other hand,
RK--MPP5, SL--MPP5, and SL--MPP7 successfully preserve positivity and
reproduce the analytic solution very well. Note that SL--MPP7 is better
at reproducing the profiles around the maxima than SL--MPP5 and
RK--MPP5, showing the importance of high-order accuracy.  As can be seen
in Figure~\ref{fig:1d_adv_sin4_error}, the order of accuracy for
RK--MPP5, SL--MPP5, and SL--MPP7 is actually close to the fifth, and the
result with SL--MPP7 is more accurate.

\subsubsection{Test-1d: Two-dimensional Advection}
\label{subsec:2D_advection}

Here, we present numerical simulations of two-dimensional advection
problems with our new schemes to test their numerical accuracy in
multi-dimensional problems. We solve a two-dimensional advection
equation
\begin{equation}
  \pdif{f(\itbold{x},t)}{t} + c_x\pdif{f(\itbold{x},t)}{x} + c_y\pdif{f(\itbold{x},t)}{y} = 0,
\end{equation}
where $c_x$ and $c_y$ are advection speed along the $x$- and $y$-direction,
respectively. We set up a two-dimensional domain with $0\le x, y\le 1$
with a periodic boundary condition and consider three initial conditions
similar to those presented in the one-dimensional tests:
a rectangular-shaped wave given by
\begin{equation}
 f(x,y,t=0) = \left\{\begin{array}{cc}
	     1 & 0.25\le x, y\le 0.75\\
		   0  & {\rm otherwise}\end{array}\right.,
\end{equation}
a sinusoidal wave
\begin{equation}
 f(x,y,t=0) = 1 + \frac{1}{2}\sin(2\pi x) + \frac{1}{2}\sin(2\pi y)
\end{equation}
and a quartic sine wave
\begin{equation}
 f(x,y,t=0) = \sin^4(4\pi x) + \sin^4(4\pi y).
\end{equation}
In all three runs, we set the advection velocities to
$(c_x,c_y)=(1,0.5)$. The simulation domain is discretized into the Cartesian
mesh grids with $N_{\rm x}$ and $N_{\rm y}$ mesh grids along the $x$- and
$y$-direction, respectively.

\begin{figure}[htpb]
 \centering
 \includegraphics[width=18cm]{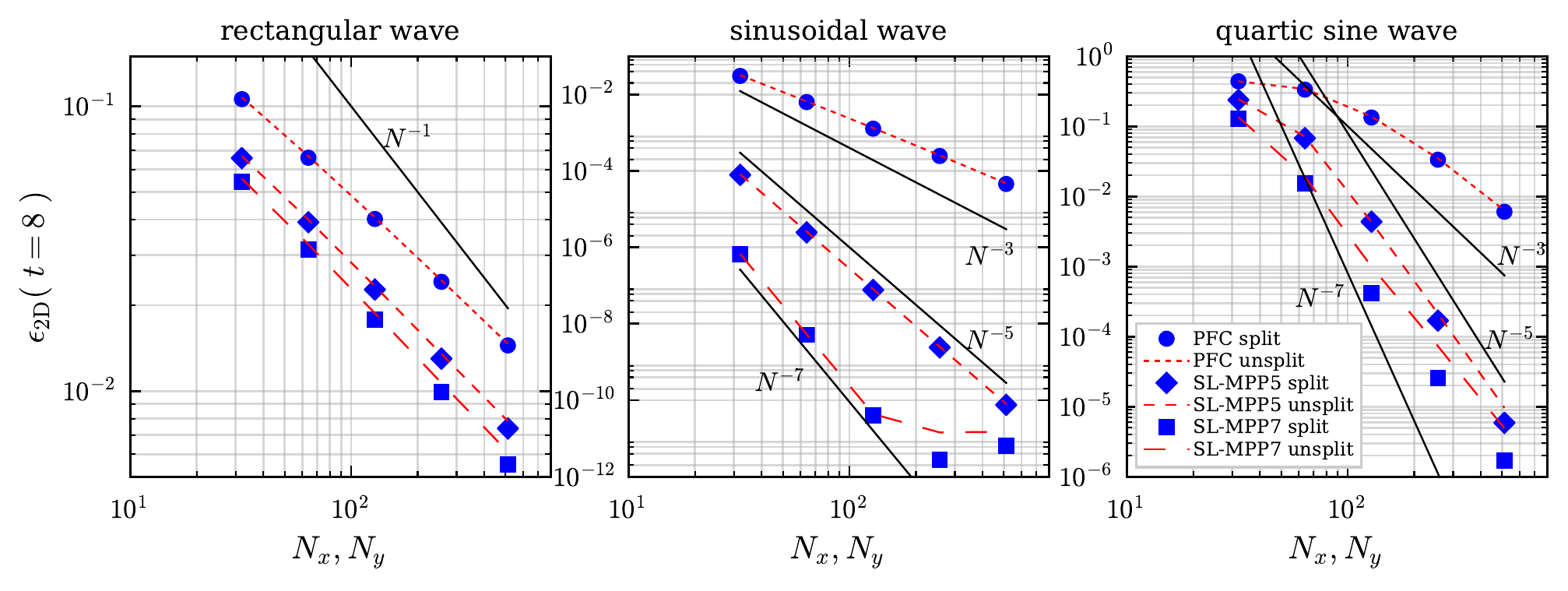}

 \figcaption{Test-1d (two-dimensional advection): relative errors are
 obtained with PFC, SL--MPP5, and SL--MPP7 in the two-dimensional
 advection problems. Points and dashed lines show the results obtained
 with the directional splitting and the unsplit methods,
 respectively. Solid lines indicate the scaling of relative errors with
 respect to the number of mesh grids as depicted in each
 panel. \label{fig:2d_advection}}
\end{figure}

We compute the relative error of numerical solutions to the analytic
solution defined by
\begin{equation}
 \epsilon_{\rm 2D} = \frac{\displaystyle \sum_{i=1}^{N_{\rm x}}
  \sum_{j=1}^{N_{\rm y}}
  \left|f_{i,j}-f(x_i, y_j,t)\right|}{\displaystyle \sum_{i=1}^{N_{\rm x}}  \sum_{j=1}^{N_{\rm y}} \left|f(x_i,y_j,t)\right|},
\end{equation}
where $f_{i,j}$ is a numerical solution at $(x,y)=(x_i,y_j)$ and $x_i$
and $y_j$ are the $x$ and $y$ coordinates of the $i$ and $j$th mesh
grids in the $x$- and $y$-direction, respectively.
Figure~\ref{fig:2d_advection} shows the relative errors of numerical
solutions obtained with PFC, SL--MPP5, and SL--MPP7 combined with the
directional splitting or the unsplit methods, in which the numbers of
mesh grids along each dimension are set to $N_{\rm x}=N_{\rm y}=32$, 64,
128, 256, and 512.

For the advection of a rectangular-shaped function, the scaling of the
relative errors is nearly first order in space irrespective of adopted
numerical schemes, while the results obtained with SL--MPP7 and SL--MPP5
are less diffusive than those obtained with PFC. On the other hand, numerical
solutions of the advection of a two-dimensional sinusoidal wave obtained
with PFC, SL--MPP5, and SL--MPP7 give the third-, fifth-, and
seventh-order scalings of the relative error, respectively, in which the
sinusoidal shape is so smooth that the MP constraint and PP limiter do
not affect the overall accuracy of the schemes. In the case of a quartic
sine function, the PP limiter compromises the numerical accuracy around
the minima of the wave, and the numerical solutions with SL--MPP5 and
SL--MPP7 exhibit nearly fifth-order accuracy, while the scaling of the
relative errors obtained with the PFC scheme is not better than the
third order. All these results are consistent with what we have in the
one-dimensional Test-1a, Test-1b, and Test-1c, indicating that our new
schemes also exhibit spatially high-order accuracies in the
multidimensional problems.

As described in section~\ref{sec:time_integration}, we do not adopt the
unsplit method for the time integration, due to its several disadvantages
against the directional splitting method. It is, however, useful to make
a comparison of numerical accuracy obtained with these two methods. The
relative errors obtained with the unsplit method combined with PFC,
SL--MPP5, and SL--MPP7 are also shown in each panel of
Figure~\ref{fig:2d_advection}.  We heuristically find that the CFL
parameter must be less than 0.1 in the runs with the unsplit method to
keep the monotonicity and positivity of the numerical solutions, and we
set $\nu=0.02$ in computing the results with the unsplit method. It is
clearly seen that the unsplit method yields almost the same level of
numerical errors as the directional splitting method. We also find that
the computational cost in computing with the unsplit method is larger
than that of the directional splitting method by $\simeq 30$\% in terms
of wall-clock time per step.

\subsection{Test-2: Vlasov--Poisson Simulation of One-dimensional Gravitational Instability and
  Collisionless Damping}

\begin{figure}[htbp]
 \centering
 \includegraphics[width=13cm]{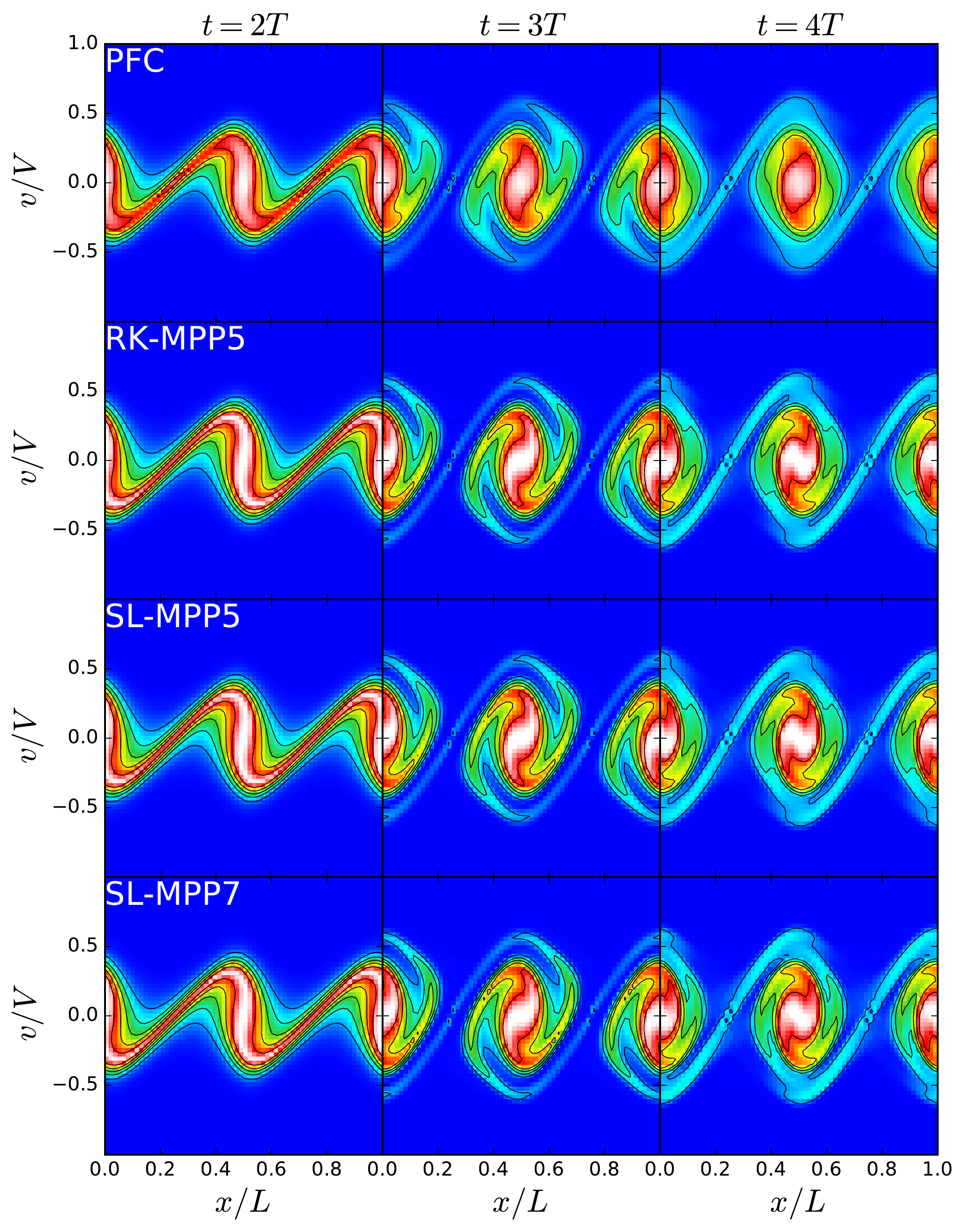}
 \figcaption{Test-2 (one-dimensional gravitational instability): Maps of
 the phase space density computed with PFC, RK--MPP5, SL--MPP5 and
 SL--MPP7 from top to bottom at $t=2T$, $3T$ and $4T$.  We set $N_{\rm
 x} = N_{\rm v} =64$.  \label{fig:maps_2d_grav_inst}}
\end{figure}

Next, we perform Vlasov--Poisson simulations of spatially
one-dimensional self-gravitating systems in two-dimensional phase space
volume. We consider the periodic simulation domain with $0\le x \le L$
and set the initial condition to be
\begin{equation}
 f(x,v,t=0) = \frac{\bar{\rho}[1+A\cos(kx)]}{(2\pi \sigma^2)^{1/2}}\exp\left(-\frac{v^2}{2\sigma^2}\right),
\end{equation}
where $\sigma$ is the velocity dispersion, $\bar{\rho}$ is the mean
density of the system, and $A$ is the amplitude of density inhomogeneity.
This distribution function yields the initial mass density as
\begin{equation}
 \rho(x) = \bar{\rho}[1+A\cos(kx)],
\end{equation}
and the density inhomogeneity develops through the gravitational
instability when the wavenumber $k$ satisfies the condition $k/k_{\rm
J}<1$, where $k_{\rm J}$ is the Jeans wavenumber defined by
\begin{equation}
 k_{\rm J} = \left(\frac{4\pi G \bar{\rho}}{\sigma^2}\right)^{1/2}.
\end{equation}
In this test, we set the wavenumber $k$ to $k=4\pi/L$ and the velocity
dispersion $\sigma$ so as to be $k/k_{\rm J}=0.5$, and the amplitude to
$A=0.01$.  The velocity space is defined over $-V \le v \le V$, where
$V=L/T$ and $T$ is the dynamical timescale defined by
$T=(G\bar{\rho})^{-1/2}$.  We adopt outflow boundary conditions in the
velocity space.  Figure~\ref{fig:maps_2d_grav_inst} shows the maps of
the phase space density computed with PFC, RK--MPP5, SL--MPP5, and
SL--MPP7 at $t=2T$, $3T$, and $4T$. The numbers of mesh grids in physical
and velocity space are set to $N_{\rm x}=N_{\rm v}=64$.  The phase space
density computed with PFC with third-order spatial accuracy is
significantly smeared especially at later epochs compared to the results
with other spatially higher-order schemes. The results with RK--MPP5 and
SL--MPP5 schemes, both of which have the spatially fifth-order accuracy,
appear almost identical to each other. SL--MPP7 is able to resolve finer
spiral-arm structure of matter distribution in the phase space located
at the turnaround position ($x\simeq 0.35$ and $x\simeq 0.65$).

\begin{figure}[htbp]
 \centering
 \includegraphics[width=13cm]{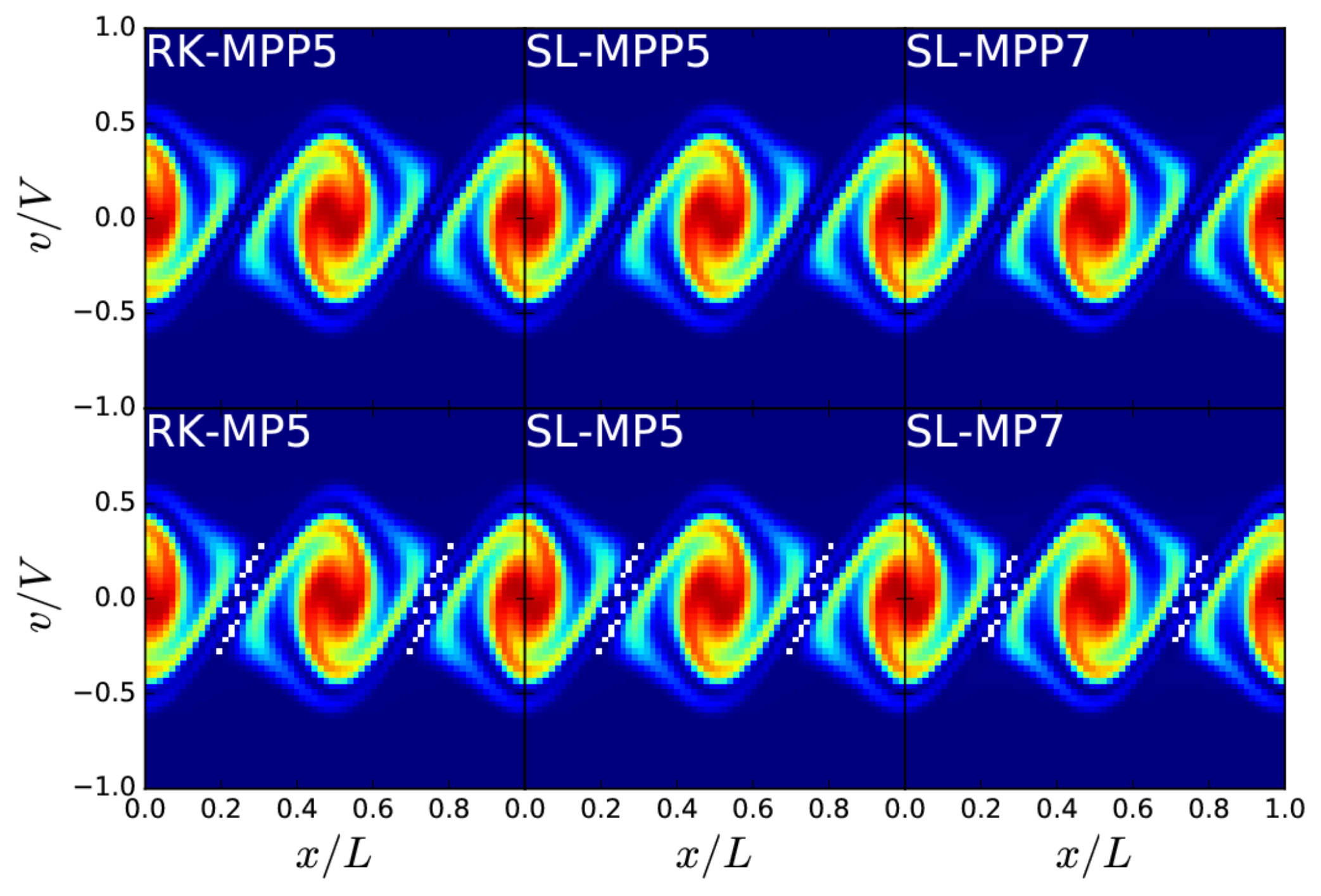}
 \figcaption{Test-2 (one-dimensional gravitational
 instability): maps of the phase space density obtained by advection
 schemes with and without the PP limiter.  White
 regions in the bottom panels have negative phase space density.
 We set $N_{\rm x} = N_{\rm v} =64$.
 \label{fig:maps_2d_grav_inst_negative}}
\end{figure}

Figure~\ref{fig:maps_2d_grav_inst_negative} compares the results
computed with RK--MPP5, SL--MPP5, and SL--MPP7 with the PP limiter (top
panels) and with RK--MP5, SL--MP5, and SL--MP7 without it (bottom
panels). The regions with negative phase space density depicted by white
dots in the bottom panels are not present in the results computed with
the PP limiter.

\begin{figure}[htbp]
 \centering
 \includegraphics[width=15cm]{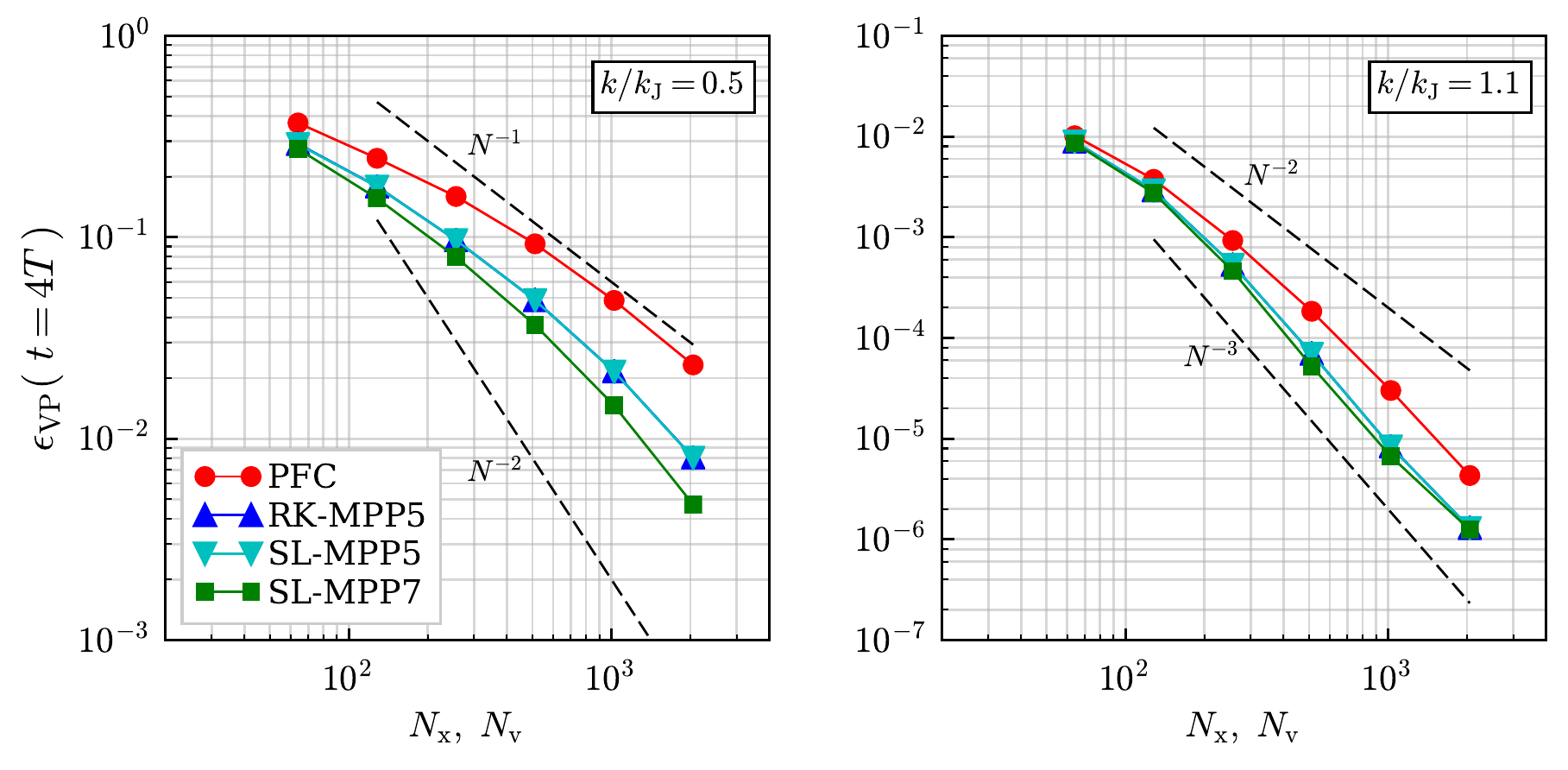}
 \figcaption{Test-2 (one-dimensional gravitational instability):
 relative errors of numerical solutions computed with the PFC, RK--MPP5,
 SL--MPP5, and SL--MPP7 schemes with respect to the reference solution as
 a function of number of mesh grids. Results with $k/k_{\rm J}=0.5$ and
 $1.1$ are shown in the left and right panels, respectively.
 \label{fig:test2_df_error}}
\end{figure}

To measure the accuracy of the numerical solutions obtained with various
schemes, we compute the following quantity:
\begin{equation}
 \epsilon_{\rm VP}(t) = \frac{\displaystyle\sum_{i=1}^{N_{\rm x}}
  \sum_{j=1}^{N_{\rm v}} |f(x_i, v_j, t)-f_{\rm ref}(x_i, v_j, t)|}{\displaystyle\sum_{i=1}^{N_{\rm x}}\sum_{j=1}^{N_{\rm v}}|f_{\rm ref}(x_i, v_j,t)|},
\end{equation}
where $f_{\rm ref}(x,v,t)$ is a reference solution to be compared
with. Since we do not have any analytical solution for this test,
we construct the reference solution $f_{\rm ref}(x,v,t)$ in the
following manner: First, we conduct runs with $N_{\rm x}=N_{\rm v}=4096$
using SL--MPP7 and apply the fifth-order spline interpolation to the
results so that the values of the numerical distribution function can be
directly compared with between different numbers of mesh grids. The left
panel of Figure~\ref{fig:test2_df_error} shows the relative errors of
the numerical solutions at $t=4T$ computed using PFC, RK--MPP5, SL--MPP5,
and SL--MPP7 with $N_{\rm x}=N_{\rm v}=64-2048$ mesh grids.
The scalings of the relative errors of our new schemes and the
PFC scheme are nearly in second and first order, respectively, and
somewhat compromised compared between their formal orders of accuracies,
because the functional shape of the distribution function is not so
smooth that the MP constraint and the PP limiter effectively work. The
accuracies of RK--MPP5, SL--MPP5, and SL--MPP7 for a given number of mesh
grids are significantly better than PFC with the same number of mesh
grids and are similar to or even better than that with four times as
many mesh grids.

We also measure the accuracies of our new schemes in the runs with
$k/k_{\rm J}=1.1$, in which the density fluctuation damps through the
collisionless damping or the free streaming. In the case with $k/k_{\rm
J}=1.1$, we set the amplitude of density fluctuation to $A=0.1$ and
extend the velocity space as $-2V\le v \le 2V$ since the velocity
dispersion is larger than the case with $k/k_{\rm J}=0.5$. We compute
the relative errors of the distribution function at $t=4T$ in the same
manner as done in the case with $k/k_{\rm J}=0.5$. The right panel of
Figure~\ref{fig:test2_df_error} shows the relative errors of numerical
solutions computed with our new schemes as well as PFC for
 $N_{\rm x}=N_{\rm v}=64-2048$ mesh grids. In this test, the
shape of the distribution function is relatively smoother than the runs
with $k/k_{\rm J}=0.5$, and the numerical solutions are not as strongly
affected by the MP constraint and the PP limiter as the runs with
$k/k_{\rm J}=0.5$. Accordingly, the scalings of the relative errors are
better than those in the case with $k/k_{\rm J}=0.5$ and are nearly
third and second order for our new schemes and PFC, respectively.  Note
that the accuracies of RK--MPP5 and SL--MPP5 are almost the same as
each other in both cases with $k/k_{\rm J}=0.5$ and 1.1, indicating that
the RK schemes and SL schemes yield almost the same accuracy in these
tests.

\subsection{Test-3: Landau Damping in Electrostatic Plasma}

We perform simulations of Landau damping in
an electrostatic plasma, in which the time evolution of the distribution
function of electrons is described by the following Vlasov equation:
\begin{equation}
 \pdif{f}{t} + v\cdot\pdif{f}{x}
  -\frac{e}{m}\pdif{\phi_{\rm s}}{x}\cdot\pdif{f}{v} = 0,
\end{equation}
where $\phi_{\rm s}$ is the electrostatic potential and $e$ and $m$ are
the charge and mass of an electron, respectively. The electrostatic
potential satisfies the Poisson equation
\begin{equation}
 \nabla^2\phi_{\rm s} = 4\pi ne \left(1 -\int_{-\infty}^{\infty} f dv\right),
\end{equation}
where $n$ is the number density of ions. In this formulation, the
distribution function is normalized such that its integration over the
velocity space gives the fraction of electron number density relative to
ions.

We consider a two-dimensional phase space with $-2\pi L \le x \le 2\pi L$
and $-6 V \le v \le 6V$, where $V\equiv \omega_{\rm pe}L$ and
$\omega_{\rm pe} = (4\pi e^2 n/m)^{1/2}$ is the electron plasma
frequency and the periodic boundary condition is imposed on the physical
space. The initial condition is given by
\begin{equation}
 f(x,v,t=0) = \frac{1}{\sqrt{2\pi\sigma^2}}
  \exp\left(-\frac{v^2}{2\sigma^2}\right)(1+A\cos kx),
\end{equation}
where the wavenumber is set to $k=1/(2L)$ and $\sigma = V$. With this
initial condition, it is expected that the electric field
damps exponentially as $\exp(-\gamma \omega_{\rm pe}t)$ with
$\gamma=0.153$.

\begin{figure}[htbp]
 \centering
 \includegraphics[width=10cm]{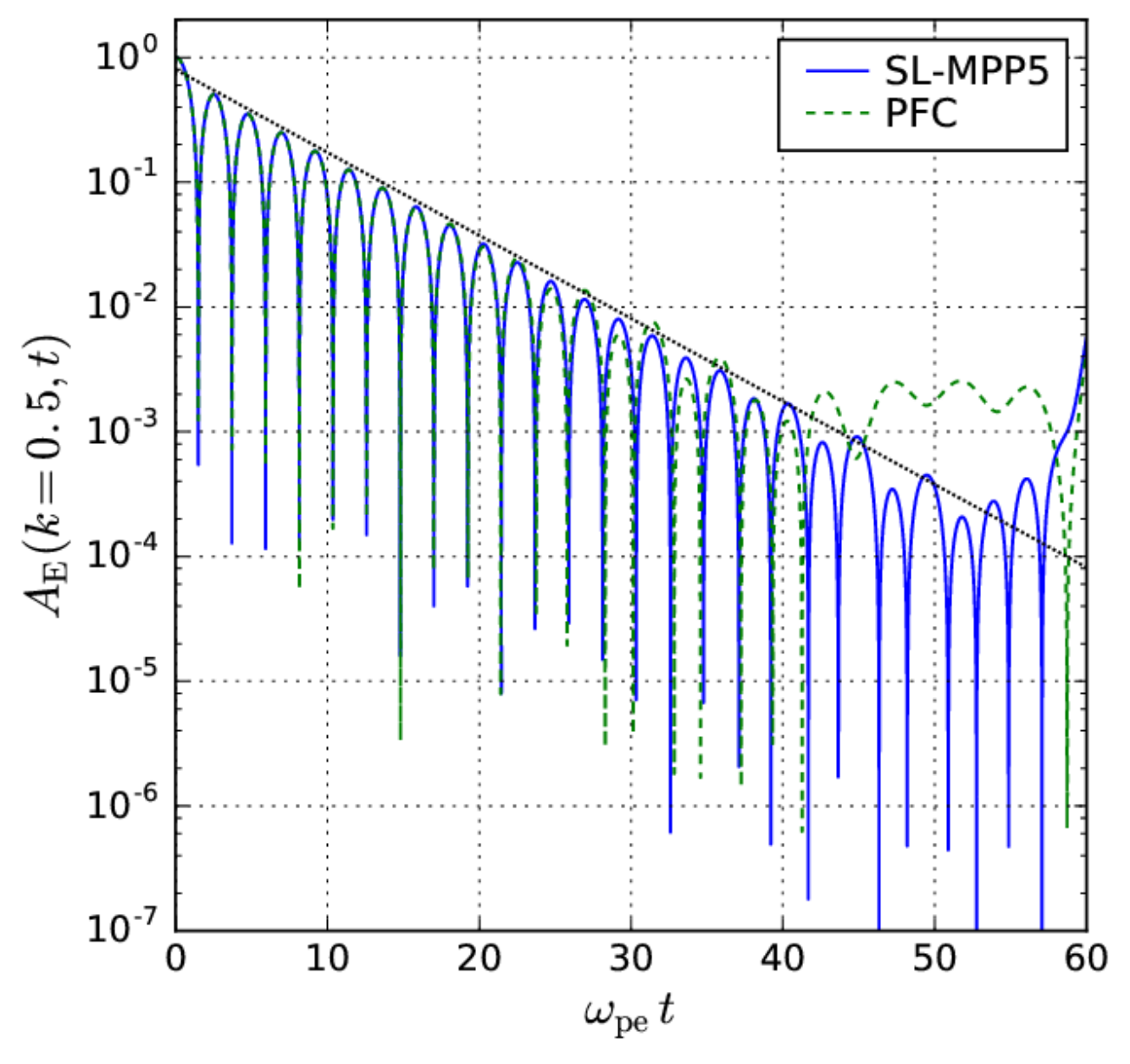}
 \figcaption{Test-3 (Landau damping in electrostatic
 plasma): time evolution of the amplitude of the electric field
 normalized at $t=0$. The dotted line indicates the theoretical
 prediction with a damping rate of $\gamma=0.153$.
 We set $N_{\rm x} = N_{\rm v} =64$.
 \label{fig:lin_LD_E_field}}
\end{figure}

First, we show results of linear Landau damping in which we set the
amplitude of the initial perturbation to $A=0.01$.  We set $N_{\rm x} =
N_{\rm v} =64$.  Figure~\ref{fig:lin_LD_E_field} shows time evolution of
the amplitude of the electric field with a mode of $k=1/(2L)$ obtained
with PFC and SL--MPP5. The results obtained with RK--MPP5 and SL--MPP7
are almost identical to that of SL--MPP5 and are not presented in this
figure.  Although the numerical solution with PFC ceases to damp at
$t\gtrsim 40 \omega_{\rm pe}^{-1}$, the one obtained with the SL--MPP5
scheme continues to damp to $t\lesssim 55 \omega_{\rm pe}^{-1}$, just
before the recurrence time $t_{\rm R}=2\pi/(k\Delta v)=67\omega_{\rm
pe}^{-1}$, indicating that SL--MPP5 is less dissipative than PFC.

\begin{figure}[htbp]
 \centering
 \includegraphics[width=15cm]{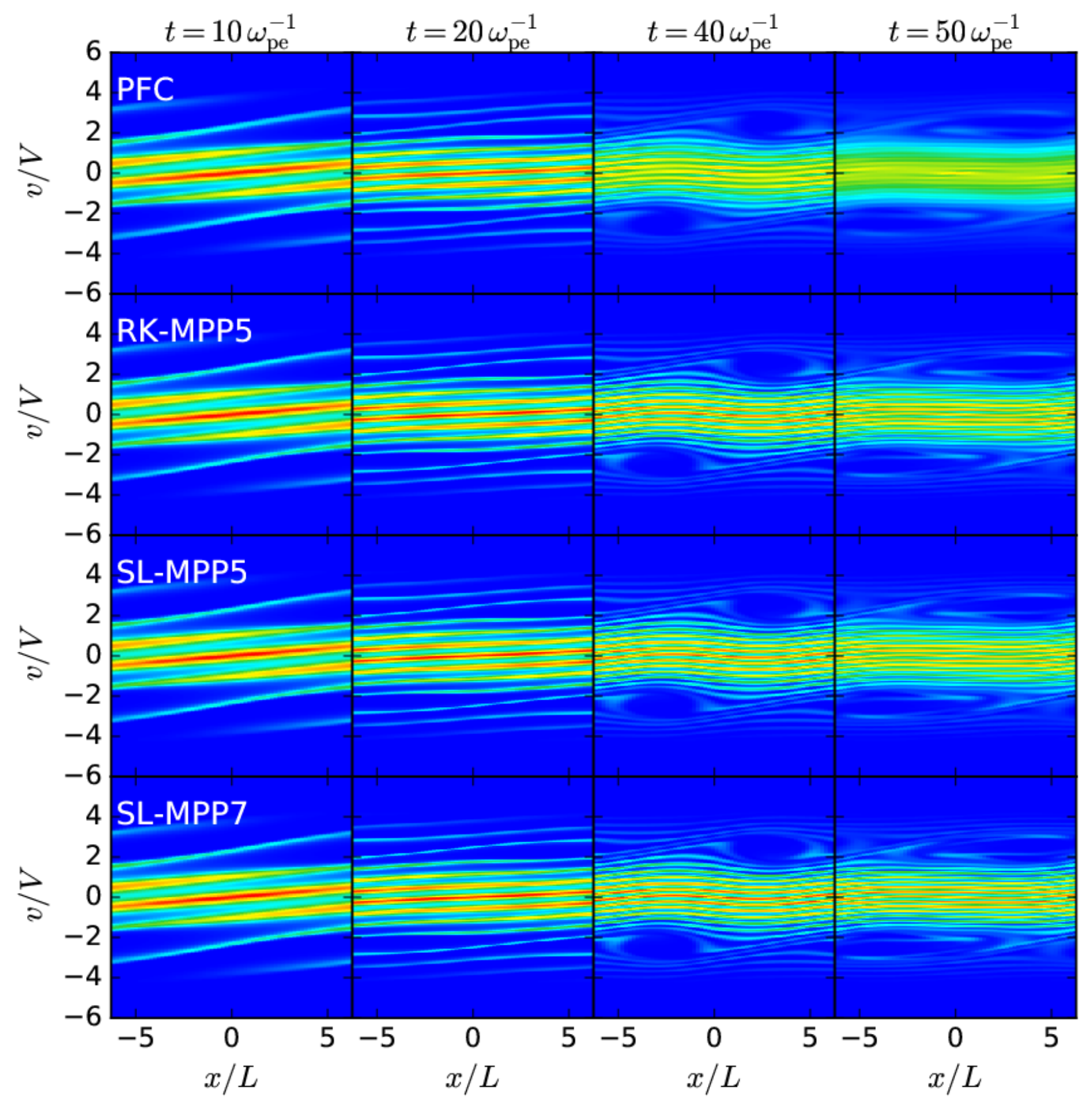}
 \figcaption{Test-3 (Landau damping in electrostatic
 plasma): maps of the phase space density of electrons in nonlinear
 Landau damping computed with PFC, RK--MPP5, SL--MPP5, and SL--MPP7
 schemes (from top to bottom) at $t=10\omega_{\rm pe}^{-1}$,
 $20\omega_{\rm pe}^{-1}$, $40\omega_{\rm pe}^{-1}$ and $50\omega_{\rm
 pe}^{-1}$ (from left to right). Numbers of mesh grids in physical and
 velocity spaces are set to $N_{\rm x}=64$ and $N_{\rm v}=256$,
 respectively. \label{fig:nonlin_LD_map}}
\end{figure}

\begin{figure}[htbp]
 \centering
 \includegraphics[width=8cm]{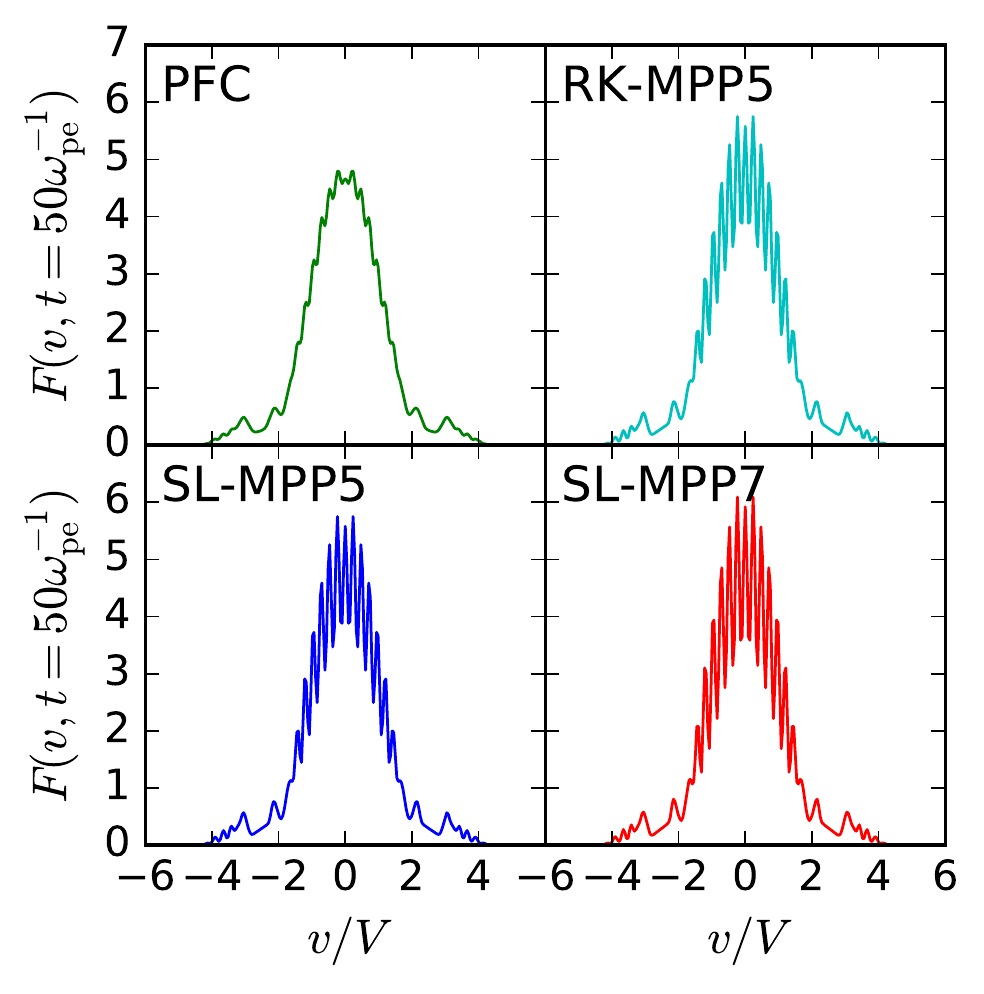}
 \figcaption{Test-3 (Landau damping in electrostatic
 plasma): velocity distribution functions in nonlinear Landau damping
 at $t=50\omega_{\rm pe}^{-1}$ computed with PFC, RK--MPP5, SL--MPP5, and
 SL--MPP7 schemes. The results are computed using the
 phase space density shown in Figure~\ref{fig:nonlin_LD_map}.
 \label{fig:nonlin_LD_vel}}
\end{figure}

Next, we perform simulations of nonlinear Landau damping with
$A=0.5$. This problem has been investigated in a number of previous
studies \citep[e.g.,][]{Klimas1987, Manfredi1997, Nakamura1999}.  In
this problem, the initial perturbations damp (in space) quickly, but a
highly stratified distribution in the phase space develops with time
through rapid phase mixing (see Figure~\ref{fig:nonlin_LD_map}). The
amplitude of the electric field initially decreases, but particle
trapping by the electric field prevents complete damping.
Figure~\ref{fig:nonlin_LD_map} shows the phase space density of
electrons at $t=10\,\omega_{\rm pe}^{-1}$, $20\,\omega_{\rm pe}^{-1}$,
$40\,\omega_{\rm pe}^{-1}$, and $50\,\omega_{\rm pe}^{-1}$ computed with
PFC, RK--MPP5, SL--MPP5, and SL--MPP7.  All the results are obtained with
$N_{\rm x} = 64$ and $N_{\rm v} = 256$.  Rapid phase space mixing
generates a number of peaks along the velocity axis. The PFC solution
exhibits a smeared distribution at $t\gtrsim 40\,\omega_{\rm pe}^{-1}$,
while those computed with other higher-order schemes preserve features
of phase mixing even at $t=50\,\omega_{\rm pe}^{-1}$. This is also seen
in Figure~\ref{fig:nonlin_LD_vel} in which we show the velocity
distribution function
\begin{equation}
 F(v,t)=\int f(x,v,t)dx
\end{equation}
at $t=50\,\omega_{\rm pe}^{-1}$. There are a number of peaks in the velocity distribution
function, which represent electrons trapped by electrostatic waves around
their phase velocities. The peaks in the distribution
function are more prominent in the results with fifth- and seventh-order schemes
than in PFC, again showing the less dissipative nature of our schemes.

\subsection{Test-4: Spherical Collapse of a Uniform-density Sphere}

Finally, we study gravitational collapse of a uniform-density
sphere. The mass and the initial radius of the sphere are $M$ and $2R$,
respectively, and the initial distribution function is given by
\begin{equation}
 f(\itbold{x}, \itbold{v}, t=0) = \frac{\rho_0}{(2\pi
  \sigma^2)^{3/2}}\exp\left(-\frac{|\itbold{v}|^2}{2\sigma^2}\right),
  \label{eq:test6_IC_DF}
\end{equation}
where $\rho_0 = M/(4\pi (2R)^3/3)$
is the initial density of the sphere
and $\sigma^2=GM/(20R)$
is the initial velocity
dispersion of the matter inside the sphere.
We set the initial virial ratio to be 0.5. The extent
of the simulation volume is $-3R \le x, y, z \le 3R$ in the physical
space and $-3V \le v_x, v_y, v_z \le 3V$, where $V$ is given by $V=R/T$
and $T=(R^3/GM)^{1/2}$ is the dynamical timescale of the system. In this
test, we adopt the out-going boundary condition in both the physical and
velocity spaces.
We perform the runs with $N_{\rm x}=64^3$ for the physical space and
$N_{\rm v}=32^3$ or $N_{\rm v}=64^3$ mesh grids for the velocity space.

\begin{figure}[htbp]
 \centering
 \includegraphics[width=13cm]{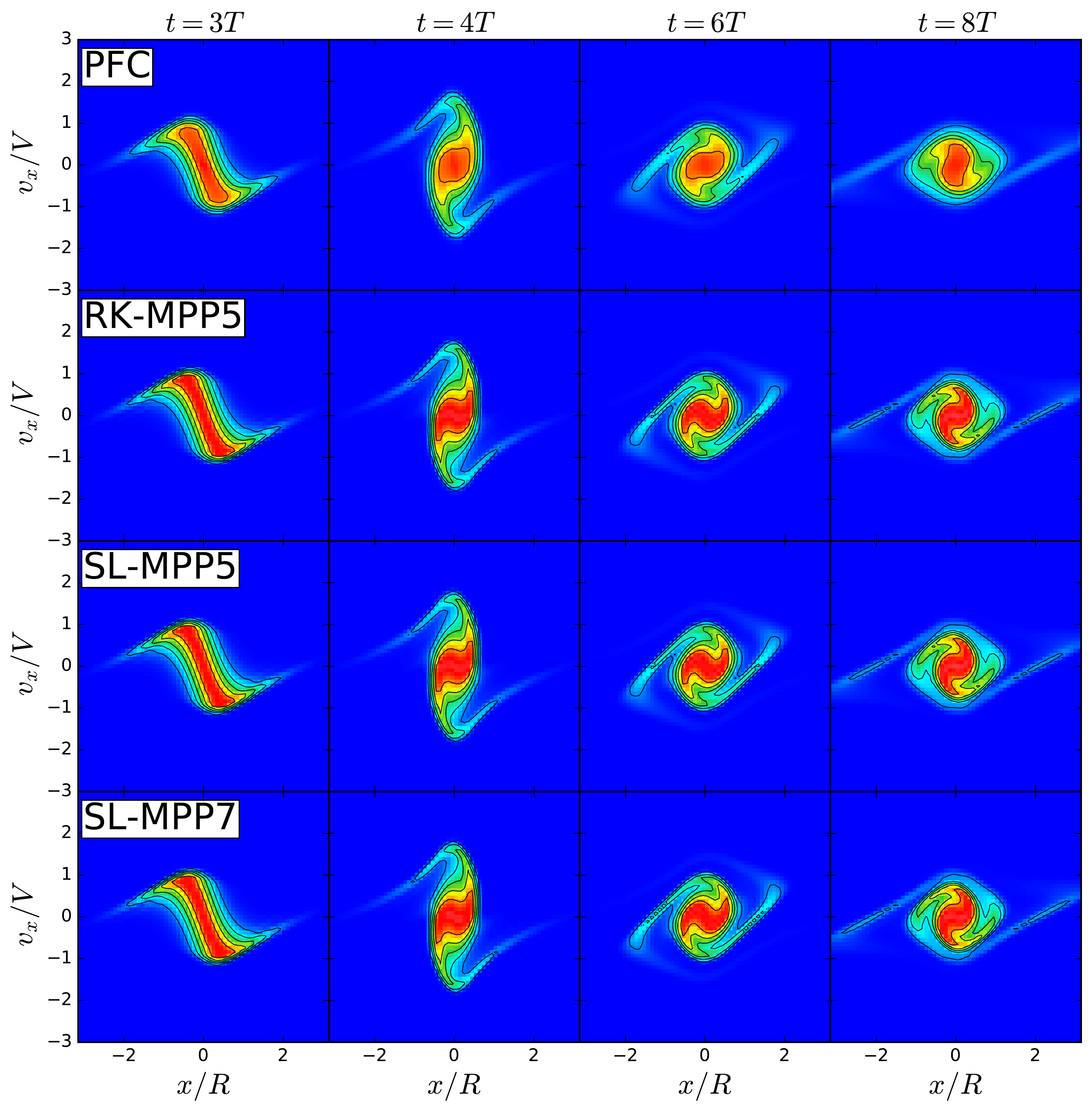}
 \figcaption{Test-4 (spherical collapse of a uniform-density sphere):
 phase space density in the ($x$,$v_x$)-plane with $y=z=0$ and $v_y =
 v_z = 0$ computed with PFC, RK--MPP5, SL--MPP5, and SL--MPP7 at $t=3T$,
 $4T$, $6T$, and $8T$. The number of mesh grids is $64^3$ in the physical
 space and $64^3$ in the velocity space. \label{fig:3d_collapse_64_64}}
\end{figure}

We plot the phase space density in the $(x,v_x)$-plane with $y=z=0$ and
$v_y=v_z=0$ at $t=3T$, $4T$, $6T$ and $8T$ in
Figure~\ref{fig:3d_collapse_64_64}.  We set $N_{\rm x} = N_{\rm v} =
64^3$.  Initially, the sphere starts to collapse toward the center and
reaches the maximum contraction around $t=4T$. Then the system expands
again, and a part of the matter passes through the center of the system
to reache the outer boundary of the physical space defined at $x=\pm
3R$, $y=\pm 3R$, and $z=\pm 3R$.  The results computed with PFC show
smeared distribution compared with those obtained with RK--MPP5,
SL--MPP5, and SL--MPP7.  We note that RK--MPP5 and SL--MPP5 yield almost
identical results, although the computational costs for the two are
significantly different (see \S~\ref{sec:computational_cost}).

\begin{figure}[htbp]
 \centering
 \includegraphics[width=13cm]{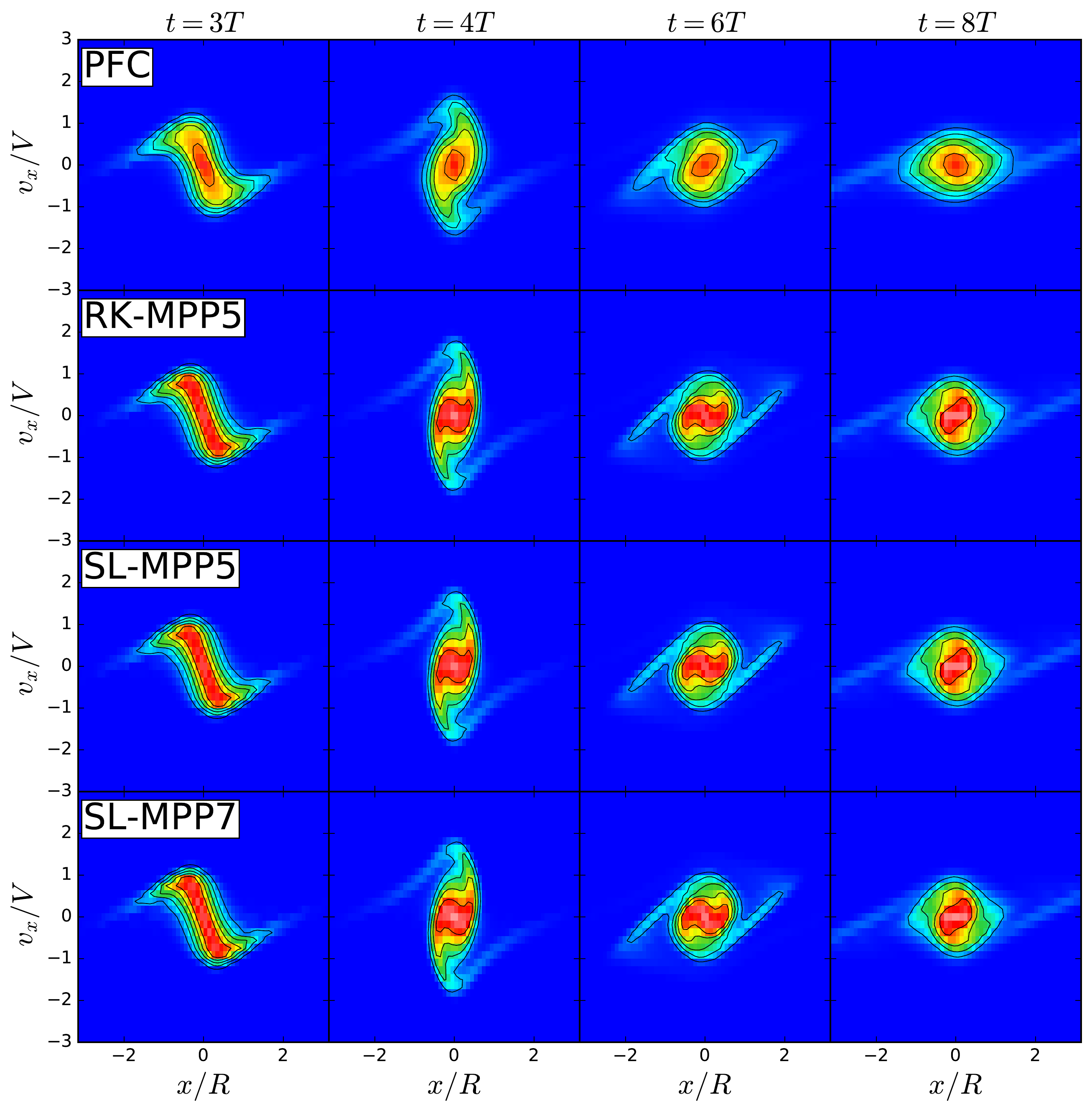}
 \figcaption{Test-4 (spherical collapse of a uniform-density
 sphere): same as Figure~\ref{fig:3d_collapse_64_64} except
 that the number of mesh grids in the velocity space is
 $32^3$. \label{fig:3d_collapse_64_32}}
\end{figure}

Figure~\ref{fig:3d_collapse_64_32} is the same plot as
Figure~\ref{fig:3d_collapse_64_64} but shows the results with
$N_{\rm v} = 32^3$.
It is interesting that the results obtained by SL--MPP7 with $N_{\rm
x}=64^3$ and $N_{\rm v}=32^3$ (bottom panels in
Figure~\ref{fig:3d_collapse_64_32}) are similar to or even better than
those obtained with PFC with $N_{\rm x}=N_{\rm v}=64^3$ (top panels in
Figure~\ref{fig:3d_collapse_64_64}), despite the eight times smaller
number of mesh grids.

\begin{figure}[htbp]
 \centering
 \includegraphics[width=13cm]{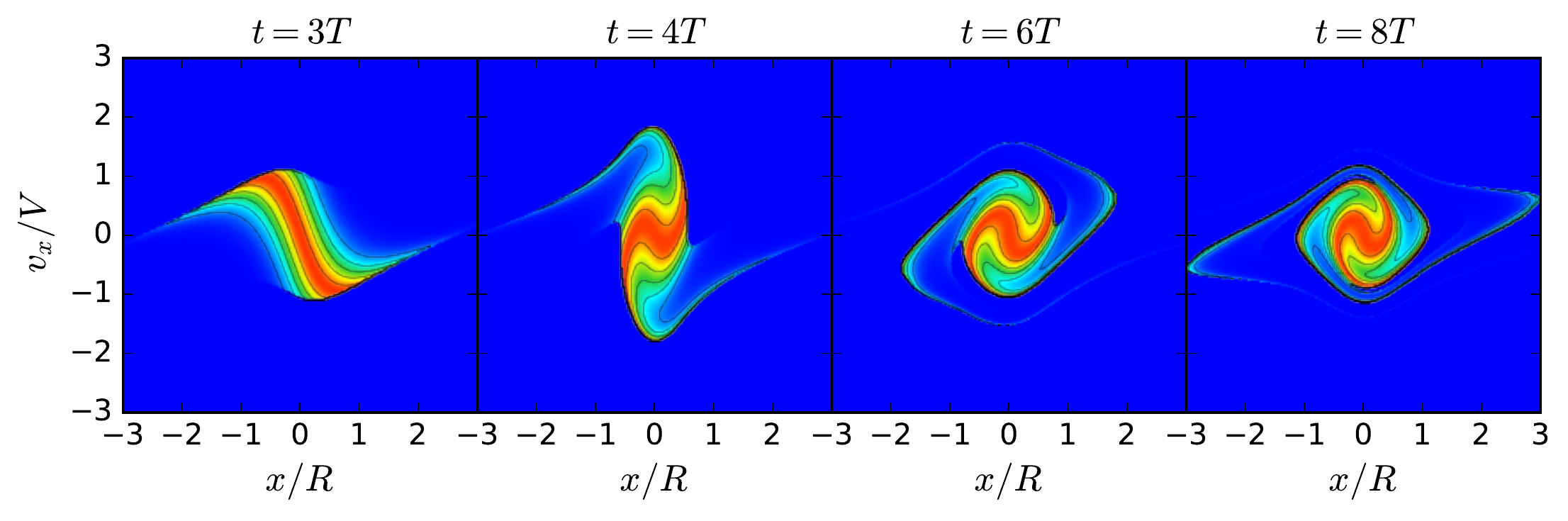}
 \figcaption{Test-4 (spherical collapse of a uniform-density
 sphere): same as Figure~\ref{fig:3d_collapse_64_64} except
 that the self-consistent field method is used instead of the
 Vlasov--Poisson simulation.\label{fig:3d_collapse_SCF}}
\end{figure}

\begin{figure}[htbp]
 \centering
 \includegraphics[width=13cm]{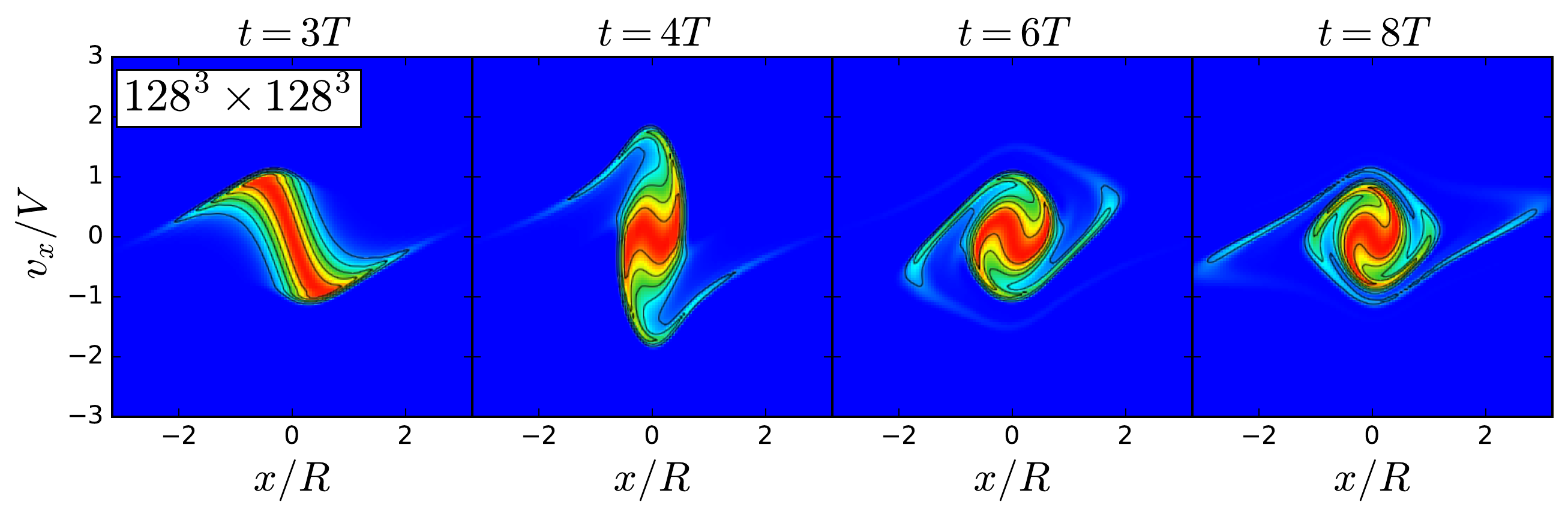}
 \figcaption{Test-4 (spherical collapse of a uniform density sphere):
 same as Figure~\ref{fig:3d_collapse_64_64} except that the number of
 mesh grids is $N_{\rm x}=128^3$ and $N_{\rm v}=128^3$. The SL--MPP5
 scheme is used for the simulation. \label{fig:3d_collapse_128_128}}
\end{figure}

For this test, we compare our new schemes with another independent
scheme.  To this end, we utilize a high-resolution $N$-body simulation
in the following manner. First, we perform an $N$-body simulation
starting from the initial condition equivalent to
Equation~(\ref{eq:test6_IC_DF}). In this $N$-body simulation, the
gravitational potential field is dumped and stored at all the time
steps. Then, we compute the phase space density at a given phase space
coordinate $(\itbold{x}(t), \itbold{v}(t))$
by tracing back the coordinate to its initial coordinate
$(\itbold{x}(0), \itbold{v}(0))$ using the previously stored
gravitational potentials. Since the phase space density is constant
along each trajectory, we can, in principle, compute the phase space
density from the initial phase space coordinate $(\itbold{x}(0),
\itbold{v}(0))$ using Equation~(\ref{eq:test6_IC_DF}). However,
trajectories in conventional $N$-body simulations are subject to rather
grainy gravitational potential compared to what it should be in the actual
physical system owing to their superparticle approximation. Thus, we
adopt the self-consistent field (SCF) method \citep{Hernquist1992,
Hozumi1997}, in which the gravitational potential is computed by
expanding the density field into a set of basis functions, and particle
trajectories are calculated for the {\it smooth} gravitational
potential.  The initial condition is constructed with $10^7$ equal-mass
particles, and the particle orbits are computed using the basis set
constructed by \citet{Clutton-Brock1972}, where the numbers of the
expansion terms are 64 and 144 in the radial and angular
directions, respectively.  Figure~\ref{fig:3d_collapse_SCF} shows the
phase space density obtained with this procedure. Comparing with Figures
\ref{fig:3d_collapse_64_64} and \ref{fig:3d_collapse_64_32} manifests
that our spatially higher-order schemes (RK--MPP5, SL--MPP5, and
SL--MPP7) yield results much closer to those obtained with the SCF
approach. It should be noted that, in practice, the SCF method is
applicable only to self-gravitating systems with a certain spatial
symmetry, such as in the present test, to which the lowest-order
functions of the basis set suffice to describe the overall structure of
the system. Also, storing the gravitational potential (or its expansion)
at all the time steps is very costly.  Therefore, the SCF approach
cannot be applied to a wide range of self-gravitating systems. For
further comparison with the results obtained by the SCF methods, we
perform a Vlasov--Poisson run with the largest numbers of mesh grids,
$N_{\rm x}=128^3$ and $N_{\rm v}=128^3$ using SL--MPP5, and show the
phase space density at $t=3T$, $4T$, $6T$ and $8T$ in the $(x,
v_x)$-plane in Figure~\ref{fig:3d_collapse_128_128}. With such high
resolution, the phase space density is consistent with that obtained
with the SCF method in detail.

\begin{figure}[htbp]
 \centering
 \includegraphics[width=13cm]{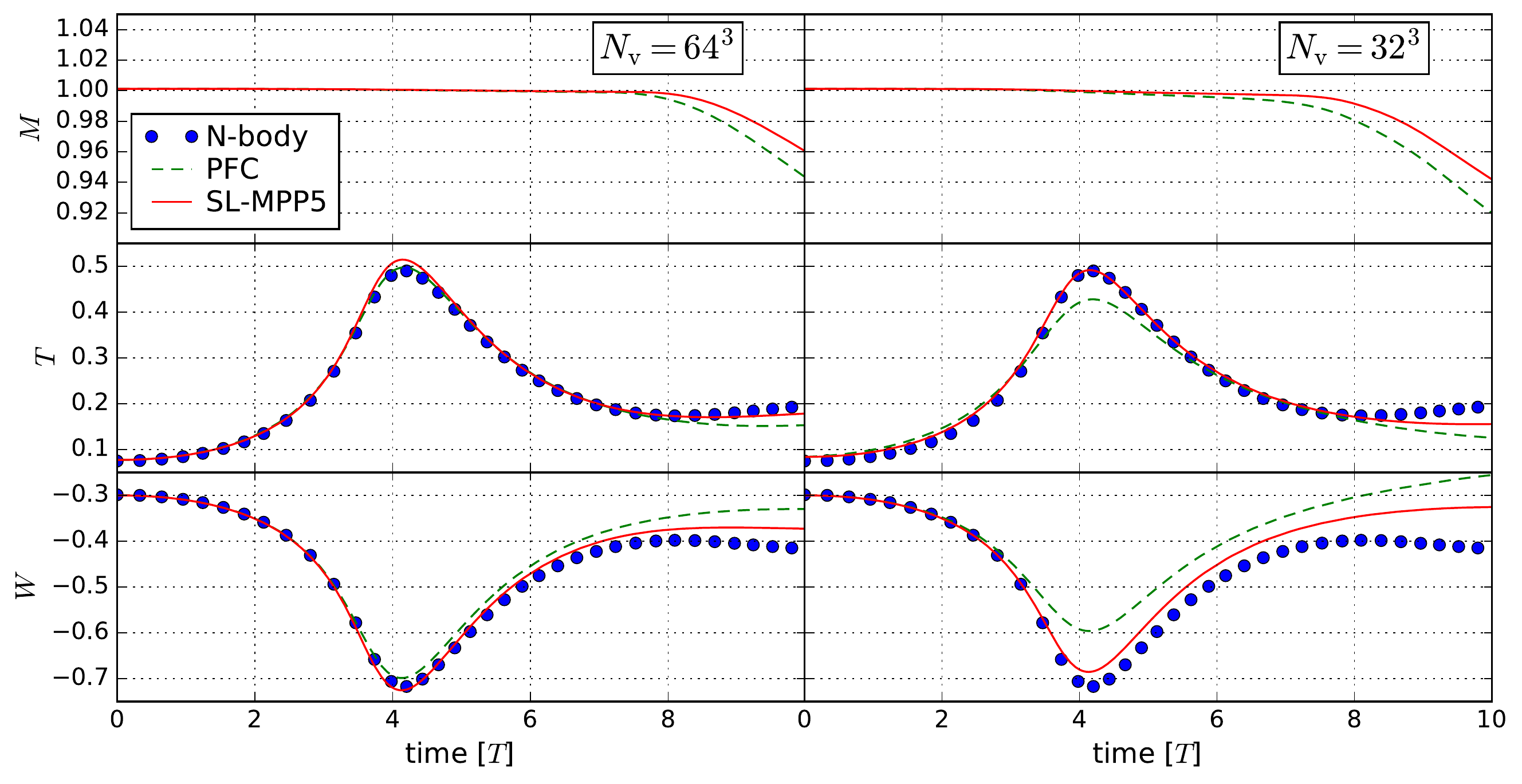}
 \figcaption{Test-4 (spherical collapse of a uniform-density sphere):
 time evolution of the total mass (top), kinetic energy (middle), and
 potential energy (bottom) in the runs with $N_{\rm v}=64^3$ (left) and
 $N_{\rm v}=32^3$ (right) computed with PFC and SL--MPP5. The kinetic
 and gravitational potential energies obtained with an $N$-body simulation
 are also shown for the comparison. \label{fig:collapse_diag}}
\end{figure}

We further examine the mass and energy conservation for this spherical
collapse test. We show the time evolution of the total mass, kinetic energy, and
gravitational potential energy of the results simulated with PFC and
SL--MPP5 in Figure~\ref{fig:collapse_diag}, where the results obtained
with $N_{\rm v}=64^3$ and $N_{\rm v}=32^3$ are shown in the left and
right panels, respectively. Note that the results simulated with
RK--MPP5 and SL--MPP7 are almost the same as those of SL--MPP5 and
hence are not shown in this figure. For comparison, we also plot the
results obtained with an $N$-body simulation, in which we employ $128^3$
particles, and adopt the Particle-Mesh (PM) method with the number of
mesh grids in solving the Poisson equation equal to that of the
Vlasov--Poisson simulation, i.e., 64 mesh grids in each dimension, in
order to realize the same spatial resolution of the gravitational field
and force. The total mass in the Vlasov simulation starts to decrease
around $t\simeq 8T$ because a part of the matter reaches the boundary of
the simulation volume at that time, as can be seen in
Figures~\ref{fig:3d_collapse_SCF} and ~\ref{fig:3d_collapse_128_128}. In
the runs with PFC, such a decrease of the total mass begins earlier and
the amount of the escaped mass is larger than that in the runs with
SL--MPP5 and other higher-order schemes owing to larger numerical
diffusion, irrespective of $N_{\rm v}$.  The most prominent difference
can be seen around the maximum contraction epoch $t\simeq 4T$, when the
kinetic and gravitational potential energies have their maximum and
minimum values, respectively.  As can be seen in
Figures~\ref{fig:3d_collapse_64_64} and \ref{fig:3d_collapse_64_32}, the
large numerical diffusion of PFC, in both the physical and velocity
spaces, smears the ``S''-shaped phase space distribution and enlarges
the physical extent of the system at $t=4T$, yielding small kinetic and
shallow gravitational potential energies seen in
Figure~\ref{fig:collapse_diag}. The overall accuracy can be improved for
PFC by simply increasing the number of mesh grids from $N_{\rm v}=32^3$
to $64^3$. On the other hand, the result of SL--MPP5 is close to the
$N$-body simulation, irrespective of $N_{\rm v}$, and in the runs with
$N_{\rm v}=64^3$ both of the results with PFC and SL--MPP5 are in good
agreement with the $N$-body simulation. This suggests that SL--MPP5 (as
well as RK--MPP5 and SL--MPP7) with $N_{\rm v}=32^3$ provides roughly
the same level of accuracy as PFC with $N_{\rm v}=64^3$ in terms of
mechanical energy.

\section{Computational Cost}
\label{sec:computational_cost}

\begin{figure}[htpb]
  \centering \includegraphics[width=13cm]{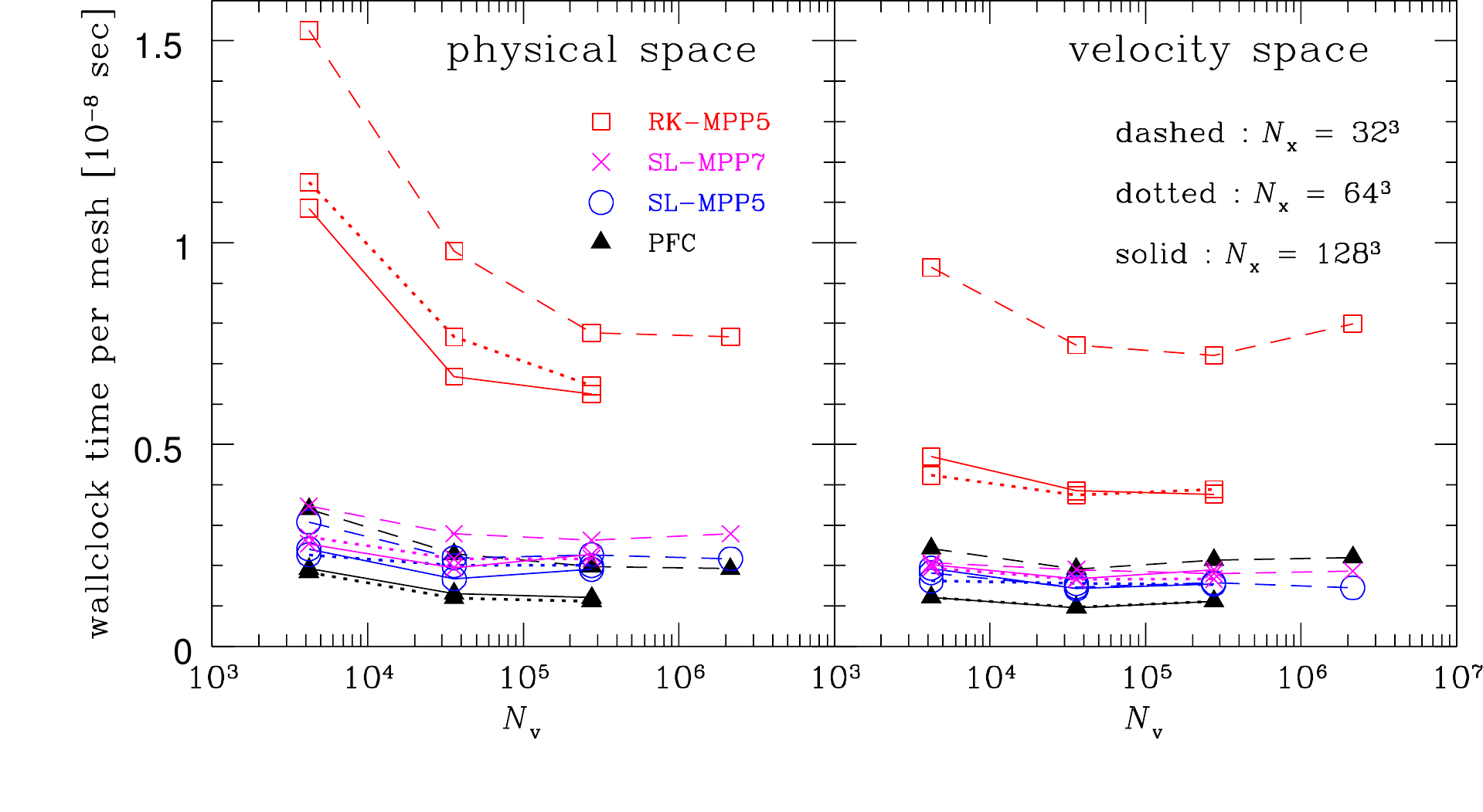}
  \figcaption{Wall-clock times per mesh grid for a single step of
  advection in the physical space (left panel) and the velocity space
  (right panel) computed with PFC (filled triangles), RK--MPP5 (open
  squares), SL--MPP5 (open circles), and SL--MPP7 (crosses). Data
  connected with solid, dotted, and dashed lines are measured in runs
  with $N_{\rm x}=32^3$, $64^3$ and $128^3$,
  respectively.\label{fig:timing}}
\end{figure}

In addition to the computational accuracy, we need to address the
computational speed of our new schemes.  We measure the wall-clock time
to advance each of Equation~(\ref{eq:adv_pos}) and (\ref{eq:adv_vel})
by a single time step using PFC, RK--MPP5, SL--MPP5, and SL--MPP7 for
various combinations of $N_{\rm x}$ and $N_{\rm v}$ ranging from $32^3$
to $128^3$ on 64 nodes of the HA-PACS system at Center for Computational
Sciences, University of Tsukuba. The largest number of mesh grids
adopted in this measurement is $(N_{\rm x}, N_{\rm v})=(128^3, 64^3)$.
Figure~\ref{fig:timing} shows the measured wall-clock time per mesh grid
in six-dimensional Vlasov--Poisson simulations of the growth of
gravitational instability from a nearly uniform-density distribution
that is identical to Test-4 in YYU13. We show the wall-clock time to
advance the advection equations in the physical space
(Equation~(\ref{eq:adv_pos})) and those in the velocity space
(Equation~(\ref{eq:adv_vel})) in the left and right panels, respectively. One
can see that RK--MPP5 is the most computationally expensive among all
the advection schemes described in this work. This is largely because it
performs three-stage RK time integration (Equation~(\ref{eq:RK}))
and computes numerical fluxes three times per step while other schemes
need only single-stage time integration (Equation~(\ref{eq:SL})).
Since the physical space is divided into subdomains and stored in
independent memory spaces, solving advection
Equation~(\ref{eq:adv_pos}) in the physical space requires internode
communications to transfer phase space densities to/from adjacent
computational nodes and thus takes a longer wall-clock time than
solving Equation~(\ref{eq:adv_vel}) in the velocity space. The
computational costs of SL--MPP5 and SL--MPP7 are significantly smaller
than that of RK--MPP5 and are almost the same as that of PFC.

\section{Summary and Conclusion}

In this work, we present a set of new spatially fifth- and seventh-order
advection schemes developed for Vlasov simulations that preserve both
of monotonicity and positivity of numerical solutions. The schemes are
based on the spatially fifth-order MP scheme
proposed by \citet{Suresh1997} combined with the three-stage TVD--RK
scheme and are extended so that they also preserve positivity by using
a flux limiter. We also devise an alternative approach to
construct MP and PP schemes by adopting
the SL time integration scheme, for the substitution of the three-stage
TVD--RK scheme. With this approach, we are able to adopt spatially
seventh- or higher-order schemes without using computationally expensive
fourth- or higher-order TVD--RK time integration schemes and hence to reduce
the computational cost.

We perform an extensive set of numerical tests to verify the improvement
of numerical solutions using our newly developed schemes. The test suite
includes one- and two-dimensional linear advection problems,
one-dimensional collisionless self-gravitating and electrostatic plasma
systems in two-dimensional phase space, and a three-dimensional
self-gravitating system in six-dimensional phase space. The results are
compared with those obtained with the spatially third-order
PFC scheme adopted in our previous work
\citep{YYU2013} and are summarized as follows.

In Test-1, we present calculations of one-dimensional linear advection
problems and confirm that numerical solutions computed with our new
schemes have better accuracy than PFC. The PP limiter
works well in every advection test, although it introduces a slight
degradation of numerical accuracy in terms of the error scaling in
combination with the TVD Runge-Kutta time integration scheme. It is found
that the SL time integration provides us with better numerical accuracy
than the TVD--RK scheme and that the combination of the PP limiter and
the SL time integration does not introduce any expense of numerical
accuracy unlike the TVD--RK time integration. It should be noted
that the spatially seventh-order RK--MPP7 exhibits undesirable features
in the advection of a rectangular-shaped function originating from
insufficient accuracy of the three-stage RK time integration scheme, and
that the numerical solutions obtained with SL--MPP7 contrastingly do
not have such features, indicating that the SL approach is crucial for
constructing schemes with spatially seventh- or higher-order
accuracy. Numerical solutions of two-dimensional linear advection
problems computed with our new semi-Lagrangian schemes are consistent
with those in the one-dimensional ones, proving that our new schemes
work well also in multidimensional problems.

In Test-2 and Test-3, we perform Vlasov--Poisson simulations of a
one-dimensional self-gravitating and electrostatic plasma systems in
two-dimensional phase space, respectively. In the self-gravitating
system, we found that the phase space density distribution simulated
with the PFC scheme is significantly suffered from numerical diffusion
compared with those simulated with our new schemes. In terms of the
$L^1$ error of the distribution function, the fifth-order schemes
(RK--MPP5 and SL--MPP5) yield almost identical results and achieve
almost the same level of numerical accuracy as the third-order scheme
with four times refined mesh grids, while the spatially seventh-order
SL--MPP7 is proved to be the most accurate scheme among all the schemes
presented in this work.  In addition, the PP limiter
also works well in the Vlasov simulations. In the simulations of
electrostatic plasma, we perform the linear and nonlinear Landau
damping problems. In the former, we found that numerical solution
computed with the third-order scheme ceases damping sufficiently before
the recurrence time, due to its large numerical diffusion. On the other
hand, the numerical solutions with our new schemes continue to damp just
before the recurrence time. As for the nonlinear Landau damping, the highly
stratified distribution formed through rapid phase mixing is well solved
with our new schemes, while it is strongly smeared in the numerical
solution computed with PFC, due to its large numerical diffusion.

Finally, in Test-4 we perform Vlasov--Poisson simulations of a fully
three-dimensional self-gravitating system in a six-dimensional phase
space, in which runs with two different sets of numerical resolution
$(N_{\rm x}, N_{\rm v}) = (64^3, 64^3)$ and $(N_{\rm x}, N_{\rm v}) =
(64^3, 32^3)$ are conducted. By comparing the distribution functions
obtained with the Vlasov--Poisson simulation with the reference solution
obtained by exploiting the SCF method, one can see that our new schemes
yield better numerical solutions than the third-order scheme, and the
advantages of these schemes are confirmed also in Vlasov simulations in
six-dimensional phase space. Comparison of the results between the runs
with two different numerical resolutions tell us that our new schemes
yield numerical solutions equivalent to those obtained by the
third-order PFC scheme with eight times refined mesh grids. We also
perform a Vlasov--Poisson simulation with $128^3$ mesh grids each for
the spatial and velocity space, $128^6$ mesh grids in total, which
successfully produce the apparently same phase space distribution as
that obtained by exploiting the SCF method.

As for the computational costs of our new schemes, RK--MPP5 is the most
expensive and requires two to three times longer wall-clock time to advance a
single time step than other schemes. This is mainly because it adopts
the three-stage TVD--RK time integration scheme and operates the
Eulerian time integration three times per step. On the other hand, the
SL schemes perform only a single Eulerian time integration per step;
hence, their computational costs are significantly less expensive than
RK--MPP5 and are nearly the same as that of PFC, indicating that the SL
schemes achieve the improvement of effective spatial resolution without
a significant increase of computational costs.

Among all the new schemes described in this paper, the spatially
seventh-order SL scheme (SL--MPP7) is the most favored one in terms of
numerical accuracy and computational cost, although the spatially
fifth-order SL scheme can be a viable option when the capacity of
internode communication is not sufficient. As a similar scheme to
SL schemes, \citet{Qiu2010} and \citet{Qiu2011} develop conservative
semi-Lagrangian WENO schemes to solve the Vlasov equation that adopt
weighted essentially non-oscillatory (WENO) reconstruction to compute
the interface values instead of the MP reconstruction adopted in our
schemes. The WENO reconstruction is, however, known to be more diffusive
and more computationally expensive compared with the MP scheme
\citep{Suresh1997}.

Since Vlasov simulations require a huge amount of memory to store the
distribution function in six-dimensional phase space volume, the size of
Vlasov simulations is mainly limited by the amount of total available
memory rather than the computational cost. Adoption of spatially higher-order
advection schemes in Vlasov simulations enables us to effectively
improve the numerical resolution without increasing the number of mesh
grids. In this sense, our new spatially higher-order advection schemes are
of crucial importance in numerically solving Vlasov equations. We
implement our new advection schemes to Vlasov--Poisson simulations that
can be applied to collisionless self-gravitating and electrostatic
plasma systems and we confirmed that our schemes really improve the
numerical accuracies of Vlasov--Poisson simulations for a given number
of mesh grids. Though we have applied our schemes only to
Vlasov--Poisson simulations, we are going to apply our schemes to
Vlasov--Maxwell simulations for magnetized plasma systems in a future
work, in which difficulties in following gyro motion around magnetic
field lines or an advection by rigid rotation in velocity space have
been reported by many authors \citep{Minoshima2011, Minoshima2013,
Minoshima2015}.

We thank Professor Shunsuke Hozumi for providing distribution function
data of Test-4 computed with his SCF code and for his valuable comments
on this work. This research is supported by MEXT as “Priority Issue on
Post-K computer” (Elucidation of the Fundamental Laws and Evolution of
the Universe) and JICFuS and uses computational resources of the K
computer provided by the RIKEN Advanced Institute for Computational
Science through the HPCI System Research project (project ID:hp160212
and hp170231). This research used computational resources of the HPCI
system provided by Oakforest--PACS through the HPCI System Research
Project (project ID:hp170123). Part of the numerical simulations are
performed using HA-PACS and COMA systems at Center for Computational
Sciences, University of Tsukuba, and Cray XC30 at Center for
Computational Astrophysics, National Astronomical Observatory of Japan.

\bibliography{ms}

\appendix

\section{Monotonicity-preserving Constraint}
\label{appendix:mp5}

The MP scheme devised by \citet{Suresh1997} computes the interface value
$f_{i+1/2}^{\rm MP}$ at a mesh boundary between the $i$th and $(i+1)$th
mesh grids with five stencil values, three in the upwind side, and two in
the downwind side in the following procedures:
\begin{enumerate}
 \item Evaluate the interpolated interface value $f_{i+1/2}^{\rm int}$
       at a mesh interface using an equation. In the case of MP5 or MP7
       schemes, for example, we adopt Equation~(\ref{eq:5th_interpolation}) or
       Equation~(\ref{eq:7th_interpolation}) in the fifth- or seventh-order
       accuracy, respectively.
 \item Evaluate $f^{\rm *} = f_i+{\rm minmod}(f_{i+1}-f_i,
       \alpha(f_i-f_{i-1}))$, where a parameter $\alpha$ controls the
       upper bound of the interface value and should hold the condition
       $\alpha\ge 2$. Throughout in this work, we set $\alpha=4$, unless
       otherwise stated.
 \item If the condition $(f_{i+1/2}^{\rm int}-f_i)(f_{i+1/2}^{\rm
       int}-f^{\rm *})<0$ holds, the resultant interface value is given
       by $f_{i+1/2}^{\rm MP} = f^{\rm int}_{i+1/2}$. If not, we should
       proceed to the next step.
 \item Compute $d_{i+1/2}^{\rm M4}\equiv{\rm minmod}(4d_i-d_{i+1},
       4d_{i+1}-d_i, d_i, d_{i+1})$ and $d_{i-1/2}^{\rm M4}$, where $d_i
       = f_{i-1}+f_{i+1}-2f_i$.
 \item Compute $f^{\rm UL} \equiv f_i+\alpha(f_i-f_{i-1})$, $f^{\rm AV} \equiv
       (f_i+f_{i+1})/2$, $f^{\rm MD} \equiv f^{\rm AV} - d^{\rm
       M4}_{i+1/2}/2$ and $f^{\rm LC} \equiv f_i + (f_i-f_{i-1})/2 +
       4d_{i-1/2}^{\rm M4}/3$
 \item Compute $f^{\rm min} \equiv \max(\min(f_i,f_{i+1}, f^{\rm MD}),
       \min(f_i,f^{\rm UL}, f^{\rm LC}))$ and $f^{\rm
       max}\equiv\min(\max(f_i, f_{i+1}, f^{\rm MD}), \max(f_i, f^{\rm
       UL}, f^{\rm LC}))$.
 \item Finally, the interface value is given by $f^{\rm MP}_{i+1/2} =
       {\rm median}(f^{\rm int}_{i+1/2}, f^{\rm min}, f^{\rm max})$.
\end{enumerate}

It should be noted that the time step constraint to keep the
monotonicity of numerical solutions depends on the parameter $\alpha$
and that the CFL parameter should satisfy the condition \citep{Suresh1997}
\begin{equation}
  \nu \le 1/(1+\alpha) = 1/5,
\end{equation}
if we set the parameter $\alpha$ to $\alpha=4$ as recommended in
\citet{Suresh1997}, although $\nu=0.4$ works well in practice. (see
Appendix~\ref{appendix:cfl} for further argument on a viable range of
the CFL parameter).

\section{CFL Condition}
\label{appendix:cfl}

\begin{figure}[htbp]
 \centering
 \includegraphics[width=12cm]{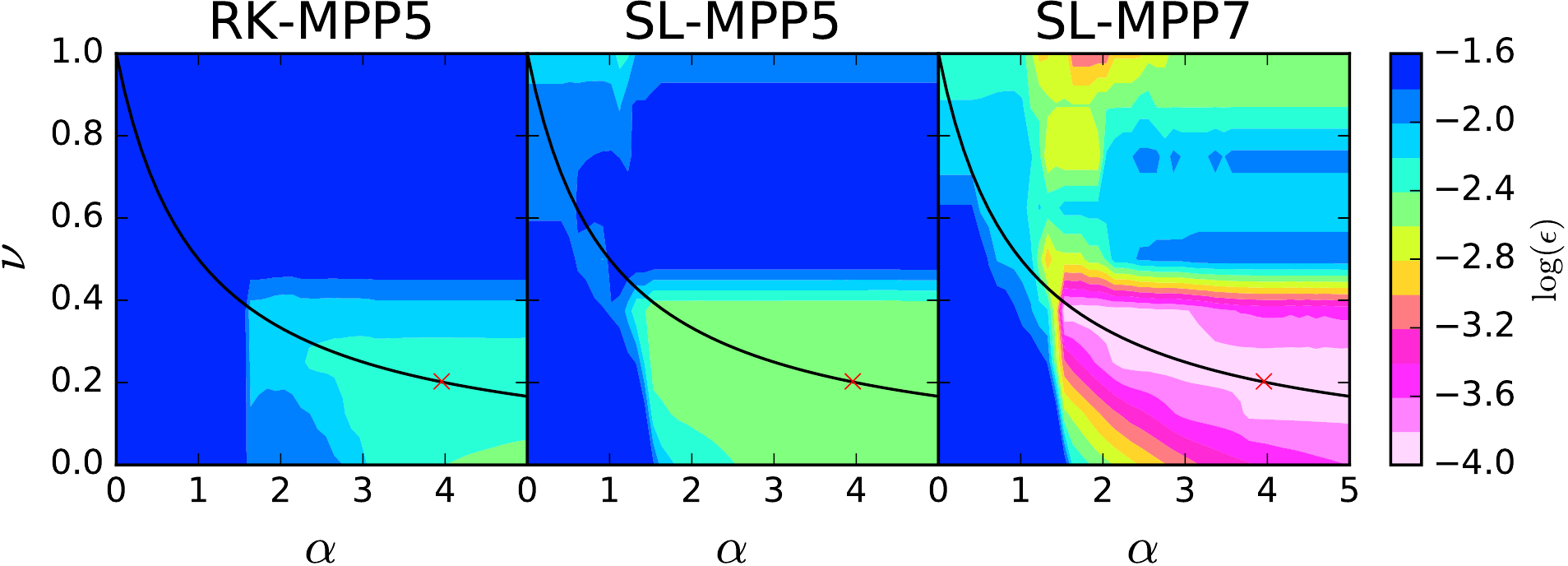}
 \figcaption{Maps of relative $L^1$ errors of Test-1c
 obtained with the RK--MPP5 (left), SL--MPP5 (middle), and
 SL--MPP7 (right) schemes for ranges of $0<\alpha<5$ and
 $0<\nu<1$. Solid curves and red crosses indicate the relation $\nu =
 1/(1+\alpha)$ and $(\alpha,\nu)=(4,0.2)$,
 respectively.\label{fig:CFL_error_map}}
\end{figure}

As noted in Appendix \ref{appendix:mp5}, the choice of $\alpha$ and
$\nu$ is rather heuristic in the MP scheme. Therefore, we also optimize
the combination of $\alpha$ and $\nu$ in our new schemes RK--MPP5,
SL--MPP5, and SL--MPP7 with the PP limiter in a heuristic manner. We
perform Test-1c (linear advection of $\sin^4(4\pi x)$) using
RK--MPP5, SL--MPP5, and SL--MPP7 with various combinations of $\alpha$
and $\nu$. The number of mesh grids is set to
128. Figure~\ref{fig:CFL_error_map} shows maps of relative $L^1$ errors
of Test-1c obtained with the three schemes for $0<\alpha<5$ and
$0<\nu<1$. Comparison of RK--MPP5 and SL--MPP5 shows that SL--MPP5
allows us to adopt a larger CFL parameter $\nu$ than RK--MPP5 to achieve
a certain accuracy. Furthermore, errors obtained with SL--MPP5 are
almost unchanged irrespective of the adopted CFL parameter as long as
$2<\alpha$ and $\nu < 0.4$, probably in virtue of its semi-Lagrangian
nature. The SL--MPP7 scheme exhibits more accurate results than the SL--MPP5
scheme for $2<\alpha$ and $\nu<0.4$, and the parameter space that
yields the most accurate results is located in the vicinity of the
relation $\nu = 1/(1+\alpha)$.

\section{The High-order Poisson solver}
\label{appendix:poisson_solver}

We derive the Green function of the discretized Poisson equation in
spatially second-, fourth-, and sixth-order accuracy.  Here, just for the
simplicity, we only consider the one-dimensional Poisson equation. The
extension to the multidimensional one is quite straightforward. Given
the discrete density field $\rho_j$ with $0\le j \le N_{\rm m}-1$, where
$N_{\rm m}$ is the number of mesh grids, approximating the Poisson
equation with the second-order discretization leads to
\begin{equation}
 \label{eq:second-order_poisson}
 \frac{\phi_{j+1}-2\phi_j+\phi_{j-1}}{\Delta x^2} = 4\pi G \rho_j.
\end{equation}
Discrete Fourier transformation (DFT) of Equation~(\ref{eq:second-order_poisson})
yields
\begin{equation}
 \hat{\phi}_k[\exp(ik\Delta x)-2+\exp(-ik\Delta x)] = 4\pi G \Delta x^2\hat{\rho}_k
\end{equation}
and $\hat{\phi}_k$ can be expressed as
\begin{equation}
 \hat{\phi}_k = - \frac{\pi G \Delta
  x^2}{\displaystyle\sin^2\left(\frac{k\Delta x}{2}\right)}\hat{\rho}_k,
\end{equation}
where the `hat' denotes the DFT of physical quantities expressed as
\begin{equation}
 \hat{q}_k = \sum_{j=0}^{N_{\rm m}-1} q_j\exp(-i kj\Delta x),
\end{equation}
and $k=2\pi n / (N_{\rm m}\Delta x)$ is the discrete wavenumber, and the
periodic boundary condition $\phi_j = \phi_{N_{\rm m}+j}$ is assumed.  The
discrete Poisson equations in the fourth- and sixth-order accuracy are
given by
\begin{equation}
 \frac{-\phi_{j+2}+16\phi_{j+1}-30\phi_{j}+16\phi_{j-1}-\phi_{j-2}}{12\Delta
  x^2} = 4\pi G \rho_j
\end{equation}
and
\begin{equation}
 \frac{2\phi_{j+3}-27\phi_{j+2}+270\phi_{j+1}-490\phi_j+270\phi_{j-1}-27\phi_{j-2}+2\phi_{j-3}}{180\Delta
  x^2} = 4\pi G \rho_j,
\end{equation}
respectively. Therefore, the corresponding Fourier-transformed
potentials can be derived as
\begin{equation}
 \hat{\phi}_k = \frac{12\pi G \Delta x^2}{\sin^2(k\Delta
  x)-16\sin^2(k\Delta x/2)}\,\hat{\rho}_k
\end{equation}
and
\begin{equation}
 \hat{\phi}_k = -\frac{180\pi G \Delta x^2}{\displaystyle
  2\sin^2\left(\frac{3k\Delta x}{2}\right)-27\sin^2(k\Delta
  x)+270\sin^2\left(\frac{k\Delta x}{2}\right)}\,\hat{\rho}_k,
\end{equation}
respectively.

Gravitational force is numerically obtained with the FDA
of the gradient of the gravitational potential. Use of
the higher-order Poisson solvers motivates us to adopt higher-order FDAs
for the gravitational force. The gradient of the gravitational potential
at $x=x_i$ is approximated as
\begin{equation}
 \nabla\phi|_{x=x_i} = \frac{\phi_{i+1}-\phi_{i-1}}{2\Delta x} + O(\Delta x^2),
\end{equation}
\begin{equation}
 \nabla\phi|_{x=x_i} =
  \frac{\phi_{i+2}-8\phi_{i+1}+8\phi_{i-1}-\phi_{i-2}}{12\Delta x} +
  O(\Delta x^4),
\end{equation}
and
\begin{equation}
 \nabla\phi|_{x=x_i} =
  \frac{\phi_{i+3}-9\phi_{i+2}+45\phi_{i+1}-45\phi_{i-1}+9\phi_{i-2}-\phi_{i-3}}{60\Delta
  x} + O(\Delta x^6)
\end{equation}
with the two-, four-, and six-point FDAs, respectively.

\begin{figure}[ht]
 \centering
 \includegraphics[width=10cm]{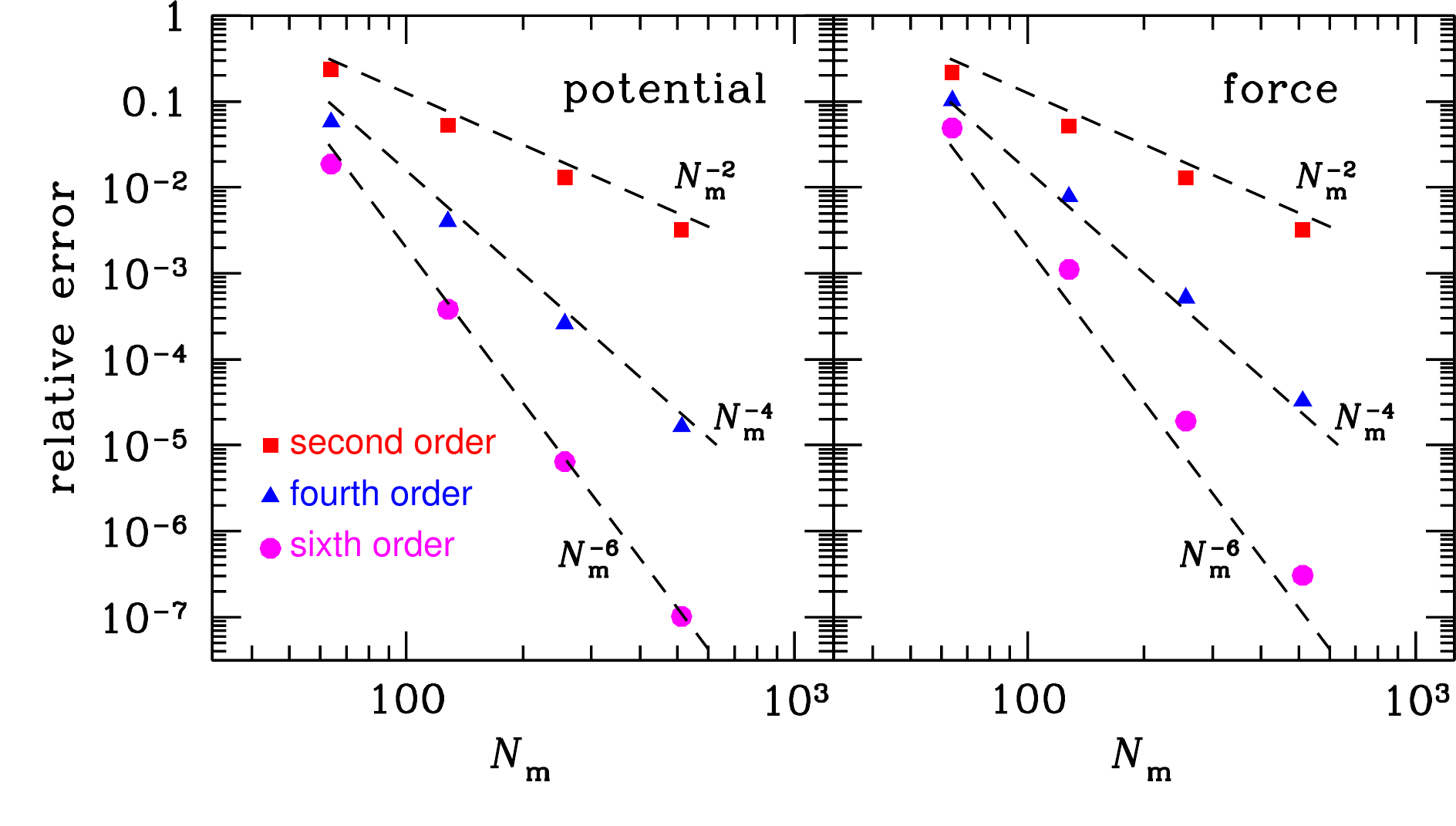}
 \figcaption{Relative errors of gravitational potentials (left panel)
 and forces (right panel) computed with spatially second-, fourth-, and
 sixth-order schemes for $N_{\rm m}=64, 128, 256$, and $512$. Dashed lines
 show the scalings of the second-, fourth-, and sixth-order accuracies from top
 to bottom.\label{fig:poisson_err}}
\end{figure}

To check the numerical accuracy of the gravitational potential and force
obtained with the numerical schemes described above, we solve the
one-dimensional Poisson equation with the periodic boundary condition
under the density field given by
\begin{equation}
 \rho(x) = \bar{\rho}\left[1 + A\sin(kx)\right],
\end{equation}
in the domain $0\le x < L$, where the wavenumber $k$ is set to
$k=32\pi/L$, and we have 16 waves in the domain. The gravitational
potential and force can be analytically solved as
\begin{equation}
 \phi(x) = -\frac{4\pi G \bar{\rho}A}{k^2}\sin(kx)
\end{equation}
and
\begin{equation}
 f(x) = -\frac{\partial \phi(x)}{\partial x} = \frac{4\pi G \bar{\rho}A}{k}\cos(kx),
\end{equation}
respectively. Figure~\ref{fig:poisson_err} shows the relative errors of
gravitational potentials and forces defined by
\begin{equation}
 \epsilon_{\rm pot} = \frac{1}{N_{\rm m}}\sum^{N_{\rm
  m}-1}_{j=0}|\phi_j-\phi(x_j)|
\end{equation}
and
\begin{equation}
 \epsilon_{\rm f} = \frac{1}{N_{\rm m}}\sum^{N_{\rm m}-1}_{j=0}|f_j-f(x_j)|,
\end{equation}
respectively. Note that the gravitational forces are computed with an FDA
scheme with the same order of accuracy as the Poisson solver, i.e., the
gravitational forces with the second-order accuracy are computed with
the gravitational potentials calculated with the second-order Poisson
solver through the second-order FDA scheme. It can be seen that both of
the gravitational potentials and forces have the expected order of the
accuracy.

\end{document}